\documentclass[9pt,twocolumn,twoside]{pnas-new}

\templatetype{pnasresearcharticle} 

\title{Automorphic Bloch theorems for hyperbolic lattices}

\author[a,1]{Joseph Maciejko}
\author[b,1]{Steven Rayan} 

\affil[a]{Department of Physics \& Theoretical Physics Institute (TPI), University of Alberta, Edmonton, Alberta T6G 2E1, Canada}
\affil[b]{Department of Mathematics and Statistics \& Centre for Quantum Topology and Its Applications (quanTA), University of Saskatchewan, Saskatoon, Saskatchewan S7N 5E6, Canada}

\leadauthor{Maciejko} 

\significancestatement{Recent experiments in circuit quantum electrodynamics and electric circuit networks have demonstrated the coherent propagation of wave-like excitations on hyperbolic lattices. The negative curvature of space that underlies such lattices invalidates the familiar Bloch theorem of ordinary solid-state physics. In this work, we prove for the first time rigorous Bloch theorems for hyperbolic lattices through the identification of appropriate periodic boundary conditions. Unlike the ordinary Bloch theorem for crystalline lattices, our hyperbolic Bloch theorem is generally nonabelian in nature, and involves infinitely-many Brillouin zones for a single lattice. Our work initiates a new chapter in band theory and establishes deep connections between condensed matter physics and algebraic geometry.}

\authorcontributions{J.M. and S.R. contributed equally to this work.}
\authordeclaration{The authors declare no conflict of interest.}
\correspondingauthor{\textsuperscript{1}To whom correspondence may be addressed. E-mail: maciejko@ualberta.ca or rayan@math.usask.ca.}


\begin{abstract}
Hyperbolic lattices are a new form of synthetic quantum matter in which particles effectively hop on a discrete tessellation of 2D hyperbolic space, a non-Euclidean space of uniform negative curvature. To describe the single-particle eigenstates and eigenenergies for hopping on such a lattice, a hyperbolic generalization of band theory was previously constructed, based on ideas from algebraic geometry. In this hyperbolic band theory, eigenstates are automorphic functions, and the Brillouin zone is a higher-dimensional torus, the Jacobian of the compactified unit cell understood as a higher-genus Riemann surface. Three important questions were left unanswered: whether a band theory can be expected to hold for a non-Euclidean lattice, where translations do not generally commute; whether a formal Bloch theorem can be rigorously established; and whether hyperbolic band theory can describe finite lattices realized in experiment. In the present work, we address all three questions simultaneously. By formulating periodic boundary conditions for finite but arbitrarily large lattices, we show that a generalized Bloch theorem can be rigorously proved, but may or may not involve higher-dimensional irreducible representations (irreps) of the nonabelian translation group, depending on the lattice geometry. Higher-dimensional irreps corrrespond to points in a moduli space of higher-rank stable holomorphic vector bundles, which further generalizes the notion of Brillouin zone beyond the Jacobian. For a large class of finite lattices, only 1D irreps appear, and the hyperbolic band theory previously developed becomes exact.
\end{abstract}

\dates{This manuscript was compiled on \today}
\doi{\url{www.pnas.org/cgi/doi/10.1073/pnas.XXXXXXXXXX}}

\begin{document}

\newcommand{\tr}{\mathop{\mathrm{tr}}}
\newcommand{\bsigma}{\boldsymbol{\sigma}}
\newcommand{\bphi}{\boldsymbol{\phi}}
\renewcommand{\Re}{\mathop{\mathrm{Re}}}
\renewcommand{\Im}{\mathop{\mathrm{Im}}}
\renewcommand{\b}[1]{{\boldsymbol{#1}}}
\newcommand{\diag}{\mathrm{diag}}
\newcommand{\sign}{\mathrm{sign}}
\newcommand{\sgn}{\mathop{\mathrm{sgn}}}
\renewcommand{\c}[1]{\mathcal{#1}}
\newcommand{\Jac}{\mathop{\mathrm{Jac}}}
\newcommand{\Mob}{\mathop{\textrm{M\"ob}}}
\newcommand{\Aut}{\mathop{\mathrm{Aut}}}
\newcommand{\g}{\mathfrak{g}}
\renewcommand{\geq}{\geqslant}
\renewcommand{\leq}{\leqslant}
\newcommand{\PBC}{\text{PBC}}

\newcommand{\halfs}{\mbox{\small{$\frac{1}{2}$}}} 
\newcommand{\Nf}{N_{\!f}}
\newcommand{\partialslash}{\partial \! \! \! /}
\newcommand{\xslash}{x \! \! \! /}
\newcommand{\yslash}{y \! \! \! /}

\newcommand{\cl}{\mathrm{cl}}
\newcommand{\mb}{\bm}
\newcommand{\ua}{\uparrow}
\newcommand{\da}{\downarrow}
\newcommand{\ra}{\rightarrow}
\newcommand{\la}{\leftarrow}
\newcommand{\mc}{\mathcal}
\newcommand{\bs}{\boldsymbol}
\newcommand{\lra}{\leftrightarrow}
\newcommand{\nn}{\nonumber}
\newcommand{\half}{{\textstyle{\frac{1}{2}}}}
\newcommand{\mf}{\mathfrak}
\newcommand{\MF}{\text{MF}}
\newcommand{\IR}{\text{IR}}
\newcommand{\UV}{\text{UV}}
\newcommand{\sech}{\mathrm{sech}}

\maketitle
\thispagestyle{firststyle}
\ifthenelse{\boolean{shortarticle}}{\ifthenelse{\boolean{singlecolumn}}{\abscontentformatted}{\abscontent}}{}

\dropcap{H}yperbolic lattices are a new form of synthetic quantum matter whereby particles propagate coherently on the sites of a regular structure that appears aperiodic from the vantage point of Euclidean geometry, but is periodic in two-dimensional (2D) hyperbolic space---a non-Euclidean space of uniform negative curvature~\cite{balazs1986}. In the circuit quantum electrodynamics (CQED) experiments of Ref.~\cite{kollar2019}, a lattice of microwave waveguide resonators is engineered in such a way that the geometry sensed by the photons as they hop on this lattice is mathematically equivalent to that of a tessellation or tiling of the Poincar\'e disk by regular hyperbolic polygons~\cite{coxeter1957}. By contrast, photons or other quantum particles hopping on a conventional crystalline lattice, like the 2D square lattice, register the geometry of a tessellation of Euclidean space. This quantum hopping can also be emulated using electric circuit networks, which have been used in recent years for the simulation of topological band structures~\cite{ningyuan2015,albert2015,lee2018} and are well-suited to the implementation of nonstandard lattice geometries. Very recently, Lenggenhager {\it et al.}~\cite{lenggenhager2021} used this technology to engineer a hyperbolic lattice and measure distinct signatures of wave propagation in hyperbolic space. Given these developments, one also anticipates implementations of hyperbolic lattices using other metamaterial platforms such as photonic crystals~\cite{lu2014,ozawa2019}. The above concrete realizations of hyperbolic lattices in the laboratory open up new vistas for the exploration of quantum mechanics in (negatively) curved space, with possibly far-reaching implications for fundamental physics in the areas of string theory~\cite{maldacena1999,gubser1998,witten1998}, quantum gravity~\cite{boyle2020,asaduzzaman2020,brower2021}, and quantum information~\cite{ryu2006,vidal2007,swingle2012,pastawski2015,breuckmann2016,lavasani2019}. In the long-wavelength limit, the Hamiltonian of a quantum particle on a hyperbolic lattice reduces to the well-known Laplace-Beltrami operator on the Poincar\'e disk~\cite{boettcher2020,bienias2021}, whose spectrum is well understood. However, when the de Broglie wavelength approaches the lattice spacing, the geometry of the tessellation strongly affects both the spectrum and wave functions~\cite{daniska2016,kollar2019b,yu2020,zhu2021,stegmaier2021}. For Euclidean lattices, periodicity leads to the formulation of Bloch band theory~\cite{SSP}, whereby energy levels and wave functions are characterized by a well-defined crystal momentum quantum number. It is not {\it a priori} obvious whether nor how band theory may be generalized to hyperbolic lattices, due to the non-Euclidean nature of their geometry.

In our previous work~\cite{maciejko2020}, a band theory of $\{4g,4g\}$ hyperbolic lattices was proposed, based on ideas from Riemann surface theory and algebraic geometry. For each integer $g>1$, the lattice admits a Fuchsian group $\Gamma$---a discrete but nonabelian group that, for negatively-curved surfaces, plays the role of the discrete, abelian translation group of Euclidean lattices. The quotient of the Poincar\'e disk by the action of $\Gamma$ produces a compact Riemann surface $\Sigma_g$ of genus $g$, which can be interpreted as a compactified unit cell. The fluxes that can be threaded through the $2g$ cycles of this compact surface form the components of a {\it hyperbolic crystal momentum} $\b{k}$ that lives in a $2g$-dimensional, toroidal {\it hyperbolic Brillouin zone} $\Jac(\Sigma_g)\cong T^{2g}$, known in algebraic geometry as the Jacobian of $\Sigma_g$. (In the 2D Euclidean case $g=1$, the compactified unit cell and its Jacobian are both 2-tori, and one recovers the familiar Brillouin zone 2-torus.) {\it Hyperbolic energy bands} can be subsequently computed, using an ansatz for Hamiltonian eigenstates as automorphic functions, i.e., functions that acquire a $\b{k}$-dependent $U(1)$ phase factor under $\Gamma$-translations. Such eigenstates were dubbed {\it hyperbolic Bloch eigenstates}, and our overarching theoretical framework, {\it hyperbolic band theory}. This framework was further developed in Ref.~\cite{boettcher2021}, where a comprehensive crystallography of hyperbolic lattices was also constructed, and applied in Ref.~\cite{ikeda2021} to the hyperbolic analog of the Hofstadter butterfly.

Reference~\cite{maciejko2020} left several important questions unanswered. First, while an infinite family of solutions to the Schr\"odinger equation was constructed, no proof was given that such solutions form a complete set. In other words, Ref.~\cite{maciejko2020} constructed hyperbolic Bloch eigenstates but did not prove a hyperbolic Bloch \emph{theorem} stating that all eigenstates are necessarily of hyperbolic Bloch form. Second, and as remarked in Ref.~\cite{boettcher2021}, the Fuchsian group $\Gamma$ is a nonabelian group which may admit higher-dimensional unitary irreducible representations (irreps\footnote{Since Wigner's theorem in quantum mechanics requires representations of symmetry groups on Hilbert space to be unitary or antiunitary, and antiunitary representations are excluded here on physical grounds, we will here use ``irrep'' to refer exclusively to {\it{unitary}} irreducible representations.}). The hyperbolic Bloch eigenstates of Ref.~\cite{maciejko2020} acquire a $U(1)$ phase factor under $\Gamma$-translations, and thus belong to a 1D irrep of $\Gamma$. It is conceivable that other irreps would appear in the spectrum of a generic hyperbolic lattice Hamiltonian. Finally, Ref.~\cite{maciejko2020} studied wave propagation on an infinite hyperbolic lattice, but did not address the problem of finite hyperbolic lattices. As in other areas of condensed matter physics, only finite lattices can be realized in the real world. Understanding whether the hyperbolic band theory of Ref.~\cite{maciejko2020} can be useful to model finite hyperbolic lattices realizable in the laboratory~\cite{kollar2019,lenggenhager2021} is a pressing question in the field. Conversely, one can ask whether the spectrum and eigenstates of a finite hyperbolic lattice have anything to do with the hyperbolic band theory of infinite lattices.

In this work, we address all three questions simultaneously. We show that as in conventional Bloch theory~\cite{SSP}, a careful consideration of finite (but arbitrarily large) hyperbolic lattices allows us to formulate a rigorous Bloch theorem predicting the possible form of all eigenstates of the Hamiltonian. In addition to the $U(1)$ hyperbolic Bloch eigenstates of Ref.~\cite{maciejko2020}, we find that eigenstates may in general transform according to $U(r)$ representations of the translation group $\Gamma$, where $r\geq 1$ is the dimension of the representation, and $U(r)$ is the group of unitary $r\times r$ matrices. Accordingly, the eigenstates of hyperbolic lattice Hamiltonians are in general subject to a \emph{nonabelian Bloch theorem}, whereby the eigenstates belonging to an $r$-fold degenerate multiplet mix under translations. The first step in obtaining these results is to formulate boundary conditions on finite hyperbolic lattices that are a suitable generalization of the periodic or Born--von K\'arm\'an boundary conditions in conventional solid-state theory~\cite{SSP}. By contrast with the Euclidean case, this is already a nontrivial problem in mathematics, which amounts to classifying all possible normal subgroups of finite index $N$ in $\Gamma$~\cite{sausset2007}. Physically, the index $N$ corresponds to the number of sites of a finite hyperbolic lattice, which we will hereafter refer to as a \emph{cluster}. We show by explicit calculations on the regular $\{8,8\}$ lattice that those eigenstates of a cluster that belong to 1D irreps of $\Gamma$ are precisely of the form predicted in Ref.~\cite{maciejko2020}, but where the allowed hyperbolic crystal momenta form a discrete set, with components valued in $2\pi\mathbb{Q}$ (mod $2\pi$). We discover that for a large fraction of clusters of size $N\leq 25$, the largest size we are able to acccess computationally, {\it all} eigenstates belong to 1D irreps and the $U(1)$ hyperbolic band theory of Ref.~\cite{maciejko2020} becomes exact. As $N$ increases, the set of allowed momenta forms an increasingly fine discretization of the hyperbolic Brillouin zone $\Jac(\Sigma_g)\cong T^{2g}$. More generally, eigenstates belonging to $r$-dimensional irreps correspond to points in a classical object in algebraic geometry: the moduli space $\c{M}(\Sigma_g,U(r))$ of stable holomorphic vector bundles of rank $r$ and first Chern class $0$ on $\Sigma_g$, which generalizes the Jacobian to arbitrary $r\geq1$. (Stability refers here to a numerical restriction on the subbundles of a given vector bundle.) A foundational result in algebro-differential geometry, the Narasimhan--Seshadri theorem~\cite{narasimhan1965}, indeed establishes that $\c{M}(\Sigma_g,U(r))$ is diffeomorphic to the space of inequivalent $r$-dimensional irreps of $\Gamma$. In contrast with Euclidean lattices, a hyperbolic lattice is thus in general characterized not by a single toroidal Brillouin zone, but by {\it multiple} Brillouin zones: the toroidal Jacobian in rank $r=1$, but also higher-rank moduli spaces with $r>1$. We demonstrate by explicit calculation on the $\{8,8\}$ lattice the existence of nonabelian Bloch eigenstates which belong to such higher-dimensional irreps, and outline a concrete proposal to realize abelian/nonabelian Bloch eigenstates experimentally using CQED and/or electric circuit networks.

\section*{Periodic boundary conditions for hyperbolic lattices}
\label{sec:PBC}

Standard proofs of Bloch's theorem in conventional solid-state physics~\cite{SSP} rely on the notion of Born--von K\'arm\'an or periodic boundary conditions (PBC). In what is often referred to as the first proof of Bloch's theorem, one begins by observing that the infinite set of all translation operators $\{T_\b{R}\}$ on Hilbert space, where $\b{R}\in\mathbb{Z}^d$ is an arbitrary Bravais lattice vector, commute mutually and thus can be simultaneously diagonalized. (We here assume for simplicity a simple hypercubic lattice in $d$ dimensions with unit lattice spacing.) The collection of all resulting eigenvalues for an arbitrary $\b{R}$ can be expressed as $\{e^{-i\b{k}\cdot\b{R}}\}$, but at this point in the proof the crystal momentum $\b{k}\in\mathbb{C}^d$ is in general complex-valued. One then considers a finite cluster with $N$ sites and imposes PBC, which requires that $T_\b{R}$ must act as the identity when $\b{R}$ is a translation spanning the length of the cluster. This forces $e^{-i\b{k}\cdot\b{R}}=1$ for such $\b{R}$, which in turn implies the key features of reciprocal space: namely that $\b{k}$ is real, periodic (i.e., defined modulo a reciprocal lattice vector), and discrete. In the second proof of Bloch's theorem, one imposes PBC from the start, expanding a trial eigenstate as a Fourier series involving a sum over momenta $\b{k}$ that obey those exact same conditions.

From a mathematical standpoint, the imposition of PBC can be given two distinct interpretations: one algebraic, and one topological (or geometrical). Algebraically, imposing PBC amounts to constructing a normal subgroup $G_\PBC$ of finite index $N$ in the translation group $G$ of the infinite lattice\footnote{A {\it normal subgroup} $H$ in a group $G$ (denoted $H\triangleleft G$) is a subgroup such that $gHg^{-1}=H$ for all $g\in G$, i.e., it is invariant under conjugation by any element of $G$. The {\it index} of $H$ is the number of distinct (right) cosets of $H$ in $G$; for a normal subgroup, left cosets and right cosets are equivalent.}. For example, a 1D chain with $N$ sites $x=1,\ldots,N$ and PBC such that the wave function satisfies $\psi(x-N)=\psi(x)$ corresponds to $G=\mathbb{Z}$, the additive group of integers, and $G_\PBC=N\mathbb{Z}=\{\ldots,-2N,-N,0,N,2N,\ldots\}$, the group of translations of $x$ that leave the wave function invariant. Although both $G$ and $G_\PBC$ are infinite, the factor group $G/G_\PBC=\mathbb{Z}/N\mathbb{Z}=\mathbb{Z}_N$ is a finite group---the residual group of translations on the finite cluster, understood as a ring with $N$ sites.

This purely algebraic construction can also be understood from the point of view of covering theory in algebraic topology~\cite{AT}. The minimal representation of the 1D infinite lattice is as the quotient space $X=\mathbb{R}/G=\mathbb{R}/\mathbb{Z}\cong S^1$, understood as a single unit cell $[0,1]\supset x$ compactified under the action of $G$. Here we identify $G\cong\pi_1(X)$ as the fundamental group of this compactified unit cell. The length-$N$ cluster can be similarly compactified by the action of $G_\PBC$ and denoted by $Y_N=\mathbb{R}/G_\PBC$, where $G_\PBC\cong\pi_1(Y_N)$. Although $X$ and $Y_N$ are both homeomorphic to the circle $S^1$, $Y_N$ is a \emph{(finite) $N$-sheeted cover} of $X$, expressed by the fact that $X\cong Y_N/\mathbb{Z}_N$ where $\mathbb{Z}_N\cong\pi_1(X)/\pi_1(Y_N)$ is the group of \emph{deck transformations} of the cover. This cover is also {\it normal} or {\it Galois}, meaning that the group of deck transformations acts transitively on the sheets of the cover (i.e., the $N$ ``copies'' of $X$ in $Y_N$). Spaces $Y_N$ whose fundamental group is a normal subgroup of $\pi_1(X)$ form normal covers of $X$. Finally, the cover in this case is also {\it abelian}, meaning that the group of deck transformations is abelian. In 2D, an example of normal subgroup of the translation group $G=\mathbb{Z}^2$ is $G_\PBC=N_x\mathbb{Z}\times N_y\mathbb{Z}$, corresponding to a finite rectangular cluster with $N=N_x\times N_y$ sites. Both $X=\mathbb{R}^2/G$ and $Y_N=\mathbb{R}^2/G_\PBC$ are homeomorphic to the 2-torus $T^2$, but the covering space $Y_N$ is a ``big'' torus tiled with $N$ square unit cells.

\subsection*{Hyperbolic clusters and normal subgroups}
\label{sec:covering}

From the algebraic point of view mentioned above, the fact that Bloch's theorem is compatible with PBC can be understood as follows. Focusing on our 1D example, the PBC condition $\psi(x-N)=\psi(x)$ is the statement that $\psi(g_\PBC^{-1}(x))=\psi(x)$ for any $g_\PBC\in G_\PBC$. Bloch's theorem is the statement that eigenstates of the Hamiltonian obey $\psi(g^{-1}(x))=\chi(g)\psi(x)$ for all $g\in G$ where $\chi(g)$ is the Bloch phase factor associated to a translation by $g$. In particular, choosing $g=g_\PBC\in G_\PBC$, we obtain the requirement on $\chi$ that $\chi(g_\PBC)=1$. Now, consider a translation by $g g_\PBC g^{-1}$ with $g_\PBC\in G_\PBC$ and $g\in G$. By the Bloch condition, we have:
\begin{align}\label{CompatPBC}
\psi(g g_\PBC^{-1} g^{-1}(x))&=\chi(g g_\PBC g^{-1})\psi(x)\nn\\
&=\chi(g)\chi(g_\PBC)\chi^{-1}(g)\psi(x)\nn\\
&=\psi(x).
\end{align}
This equality is ensured if $g g_\PBC g^{-1}=g'_\PBC\in G_\PBC$, which is the condition that $G_\PBC$ be a normal subgroup of $G$.

\begin{figure}[t]
\centering\includegraphics[width=0.5\columnwidth]{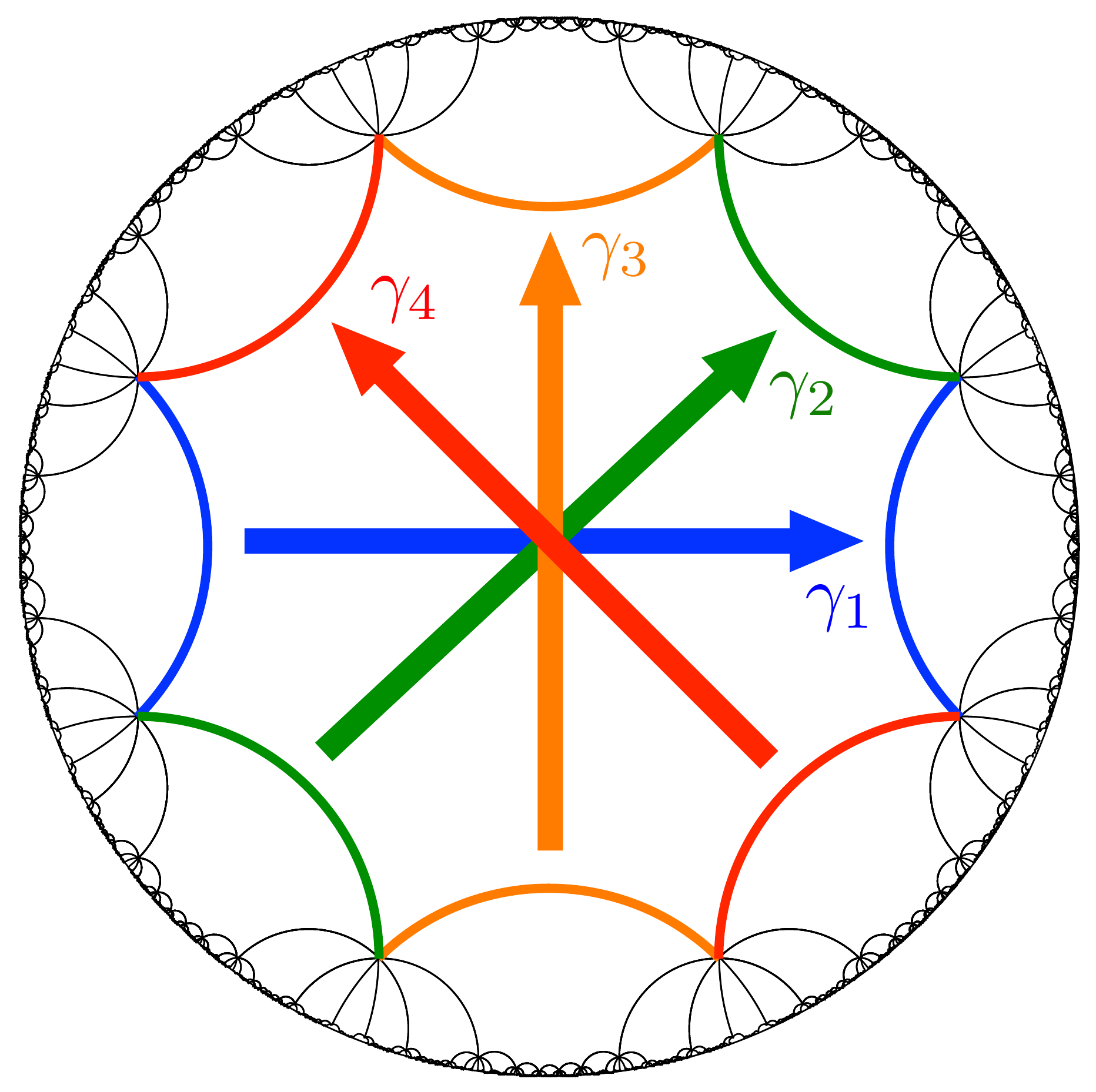}
 \caption{The Bolza or regular $\{8,8\}$ lattice and the side-pairing generators $\gamma_1,\gamma_2,\gamma_3,\gamma_4$ of its Fuchsian group $\Gamma$.}
  \label{fig:bolza}
\end{figure}

Before transposing these ideas to the hyperbolic context, we give a brief introduction to hyperbolic lattices; a more detailed discussion can be found in Refs.~\cite{maciejko2020,boettcher2021}. For our purposes, we will define a hyperbolic lattice with translation group $\Gamma$ as the discrete, infinite set of points $\{z_\gamma=\gamma(0),\gamma\in\Gamma\}$ in 2D hyperbolic space $\mathbb{H}$, represented for concreteness by the Poincar\'e disk model~\cite{[{For an introduction to hyperbolic geometry accessible to physicists, see, e.g., }]balazs1986}. This is the unit disk $\{|z|<1\}$ equipped with the Poincar\'e metric,
\begin{align}
ds^2=\frac{4(dx^2+dy^2)}{(1-|z|^2)^2},
\end{align}
in line-element form, with $z=x+iy$. With this metric, $\mathbb{H}$ is a 2D noncompact manifold with uniform negative curvature, whose full group of (orientation-preserving) isometries is the nonabelian group $PSU(1,1)$ of M\"obius transformations:
\begin{align}\label{mobius}
z\mapsto\gamma(z)=\frac{\alpha z+\beta}{\beta^*z+\alpha^*},
\hspace{5mm}\alpha,\beta\in\mathbb{C},
\hspace{5mm}|\alpha|^2-|\beta|^2=1.
\end{align}
The lattice translation group $\Gamma$ is a Fuchsian group~\cite{Katok}, i.e., an infinite discrete subgroup of $PSU(1,1)$, which we further require to be strictly hyperbolic and co-compact. These latter two conditions ensure (1) that elements in $\Gamma$ have no fixed points when acting on $\mathbb{H}$, and can thus be interpreted as translations; (2) that the unit cell is a compact region in $\mathbb{H}$. For the rest of the paper, and unless otherwise specified, we will focus for simplicity on the regular $\{8,8\}$ lattice, for which $\Gamma$ can be given the presentation:
\begin{align}\label{GammaPresentation}
\Gamma=\langle\gamma_1,\gamma_2,\gamma_3,\gamma_4:\gamma_1\gamma_2^{-1}\gamma_3\gamma_4^{-1}\gamma_1^{-1}\gamma_2\gamma_3^{-1}\gamma_4\rangle,
\end{align}
with one relation among four generators $\gamma_j$, $j=1,\ldots,4$, whose explicit $PSU(1,1)$ representation can be given as
\begin{align}\label{alphabeta}
\alpha_j=1+\sqrt{2},\hspace{5mm}\beta_j=(2+\sqrt{2})\sqrt{\sqrt{2}-1}e^{i(j-1)\pi/4},
\end{align}
in the notation of Eq.~(\ref{mobius}). The discrete set of points $\{z_\gamma\}$ is the collection of geometric centers of the hyperbolic regular octagons that tile $\mathbb{H}$ under the action of $\Gamma$; hereafter we denote by $\c{D}$ the octagon centered at $z=0$. $\c{D}$ or any of its copies is a fundamental domain for the action of $\Gamma$ on $\mathbb{H}$ (Fig.~\ref{fig:bolza}).

We now return to our discussion of the Bloch condition. In analogy with the Euclidean case reviewed above, Ref.~\cite{sausset2007} proposed that a choice of PBC for a hyperbolic lattice with Fuchsian group $\Gamma$ corresponds to a choice of normal subgroup $\Gamma_\PBC$. The first result of our work is the simple observation that such a choice of PBC is compatible with the automorphic Bloch condition\footnote{By contrast with the convention $\psi(\gamma(z))=\chi(\gamma)\psi(z)$ used in the analytic number-theory literature (e.g., Ref.~\cite{Venkov}) and in our previous work, we utilize here a convention more standard in physics~\cite{Tinkham}, and more natural from a representation-theoretic standpoint.},
\begin{align}\label{AutomorphicBloch}
\psi(\gamma^{-1}(z))=\chi(\gamma)\psi(z),
\end{align}
introduced in Ref.~\cite{maciejko2020}. Indeed, the derivation in the paragraph surrounding Eq.~(\ref{CompatPBC}) above remains entirely valid if we replace $x\in\mathbb{R}$ by $z\in\mathbb{H}$, $G$ by $\Gamma$ (as well as the group element $g$ by $\gamma\in\Gamma$), and $G_\PBC$ by $\Gamma_\PBC$ (as well as the group element $g_\PBC$ by $\gamma_\PBC\in\Gamma_\PBC$). Importantly, note that nothing in the derivation requires $G$ or $G_\PBC$ to be abelian. Furthermore, and anticipating our introduction of a nonabelian Bloch theorem later, the compatibility holds even if the factor of automorphy $\chi$ is generalized to higher-dimensional, nonabelian representations of $\Gamma$.

\begin{figure}[t]
\centering\includegraphics[width=0.8\columnwidth]{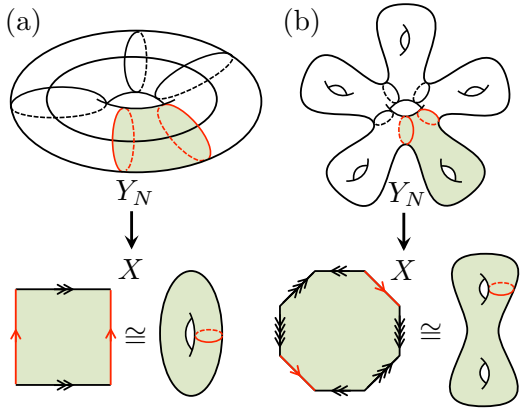}
 \caption{Examples of covering maps $Y_N\rightarrow X$ for $N=5$ with group of deck transformations $\mathbb{Z}_5$. (a) Euclidean case: a toroidal cluster covers a toroidal unit cell; (b) hyperbolic case: a genus-6 cluster covers a genus-2 unit cell. In both cases, the red loops in $Y_N$ are identified in the quotient $Y_N/\mathbb{Z}_5\cong X$.}
  \label{fig:galois}
\end{figure}

If we further consider normal subgroups $\Gamma_\PBC$ of finite index $N$ in $\Gamma$, as done implicitly in Ref.~\cite{sausset2007}, the factor group $\Gamma/\Gamma_\PBC$ is a finite group of order $N$. To each such normal subgroup corresponds a finite portion of hyperbolic lattice, or {\it cluster}, with $N$ sites. From the point of view of covering theory, the minimal representation of the infinite $\{8,8\}$ lattice is the quotient $X=\mathbb{H}/\Gamma$, a compact Riemann surface of genus 2 known as the Bolza surface~\cite{maciejko2020,bolza1887}. The Fuchsian group $\Gamma$ is thus isomorphic to the fundamental group of a genus-2 surface, which can indeed be given the presentation (\ref{GammaPresentation}). In general, if $Y_N$ is an $N$-sheeted cover of a topological space $X$, then the Euler characteristic of $Y_N$ is $N$ times that of $X$~\cite{AT}. If $X$ is a surface of genus $g$, $Y_N=\mathbb{H}/\Gamma_\PBC$ will be a surface of genus $h$ given by $2-2h=N(2-2g)$ and thus $h=N(g-1)+1$. In the 2D Euclidean case reviewed above, the compactified unit cell $X$ has $g=1$ and thus the covering space $Y_N$ (the ``big'' torus) also has $h=1$ for any $N$ [Fig.~\ref{fig:galois}(a)]. By contrast, a hyperbolic PBC cluster is necessarily a higher-genus surface, with genus $h$ that {\it grows with the size of the system} [Fig.~\ref{fig:galois}(b)]. For the $\{8,8\}$ lattice considered here, $g=2$, and thus a PBC cluster with $N$ sites has genus
\begin{align}\label{RH}
h=N+1.
\end{align}
In the context of algebraic geometry, the covering map $Y_N\rightarrow X$ is a holomorphic map $\Sigma_h\rightarrow\Sigma_g$ between Riemann surfaces that preserves the Poincar\'e metric, and the relation between $h$ and $g$ is known as the Riemann--Hurwitz formula (in the simplest case of a finite, Galois, unramified covering)~\cite{Khovanskii}. Covering theory implies that $\Gamma_\PBC$ is isomorphic to the fundamental group of a genus-$h$ surface, which can be given a finite presentation analogous to that of Eq.~(\ref{GammaPresentation}), but with $2h$ generators and one relation [see Eq.~(\ref{pi1Sigmah})]. Finally, in analogy with the Euclidean case, we interpret the factor group $\Gamma/\Gamma_\PBC$ as a finite group of translations on the cluster. Constructing a Bloch theory for finite hyperbolic clusters with PBC thus amounts to studying the representations on Hilbert space of this finite group, which we will do in {\it Fuchsian translation symmetry in finite size}.

\subsection*{The low-index normal subgroups procedure}
\label{sec:LINS}

In the Euclidean case, all subgroups of the abelian group $\mathbb{Z}^2$ are normal, and are easily enumerated. By contrast, and as noticed in Ref.~\cite{sausset2007}, the enumeration of normal subgroups of finite index in a nonabelian Fuchsian group is a nontrivial mathematical problem. To simplify the problem, one can further impose that normal subgroups be {\it torsion-free}. Torsion elements\footnote{A {\it torsion element} in a group is an element $a$ such that $a^m$ equals the identity in the group for some power $m\in\mathbb{Z}^+$.} in a Fuchsian group correspond to elliptic isometries, i.e., transformations in the same conjugacy class as a rotation $z\mapsto e^{i\alpha}z$ by angle $\alpha$ about the center of the Poincar\'e disk. (If $\Gamma_\PBC$ contained elliptic elements, the cover $Y_N=\mathbb{H}/\Gamma_\PBC$ would not be a smooth Riemann surface but an orbifold, with conical singularities.) Proofs of existence and a discussion of examples of torsion-free normal subgroups for certain Fuchsian groups can be found in the mathematical literature~\cite{hoare1971,feuer1971,edmonds1982,burns1983,kulkarni1984}. These studies consider general Fuchsian groups containing hyperbolic, elliptic, and parabolic elements. In the case of interest to us, the translation group $\Gamma$ is strictly hyperbolic, thus it and all its subgroups are necessarily torsion free. Indeed, as discussed earlier, $\pi_1(\Sigma_h)\cong\Gamma_\PBC$ can be given the presentation
\begin{align}\label{pi1Sigmah}
\pi_1(\Sigma_h)=\langle a_1,b_1,\ldots,a_h,b_h:[a_1,b_1]\cdots[a_h,b_h]\rangle,
\end{align}
where $[a,b]=aba^{-1}b^{-1}$ is the commutator of two elements in the group. This presentation does not contain any torsion.

Although for any cluster of size $N$, the associated normal subgroup $\Gamma_\PBC$ is necessarily isomorphic to (\ref{pi1Sigmah}) and thus completely known as an abstract group, in practice one needs to know how its generators $a_1,b_1,\ldots,a_h,b_h$ are expressed in terms of the original generators $\gamma_1,\ldots,\gamma_4$ of $\Gamma$, i.e., one needs to know the precise isomorphism $\pi_1(\Sigma_h)\rightarrow\Gamma_\PBC$ with $\Gamma_\PBC$ considered as a subgroup of $\Gamma$. Only then can one determine which $N$ sites $z_\gamma$ of the infinite lattice are included in a given cluster, how the boundary sites of the cluster are to be identified under PBC, and how the group $\Gamma/\Gamma_\PBC$ of residual translations acts on the sites of the cluster. In other words, for a given index $N$, there are many distinct normal subgroups of $\Gamma$, although they are all isomorphic from an abstract point of view. From a topological or geometric standpoint, there are many ways to ``wrap'' a cluster of $N$ hyperbolic unit cells into a genus-$(N+1)$ surface.

While analytical approaches appear to be of limited use for our problem~\cite{hoare1971,feuer1971,edmonds1982,burns1983,kulkarni1984}, methods in computational group theory exist that allow for the systematic enumeration of normal subgroups of a given index in a finitely presented group~\cite{sims1994}. One such method, the {\it low-index subgroups algorithm}~\cite{dietze1974}, is based on a systematic enumeration of all cosets of a given finite-index subgroup using the so-called Todd--Coxeter coset enumeration procedure~\cite{todd1936}. For normal subgroups, our prime focus here, the method was given more efficient adaptations and implementations by Conder and Dobcs\'anyi~\cite{conder2005}, and Firth~\cite{FirthThesis} in collaboration with D. Holt. Here we will use a freely available implementation of the Firth--Holt algorithm written for the computational discrete algebra system GAP~\cite{GAP4} by F. Rober~\cite{LINS}, and referred to hereafter as LINS (Low-Index Normal Subgroups)\footnote{As a test of the LINS package, we have verified that it correctly reproduces the number of normal subgroups of another Fuchsian group, the modular group $PSL(2,\mathbb{Z})$, which has been computed for indices up to 66 in Ref.~\cite{newman1967}.}.

LINS takes as sole input the presentation (\ref{GammaPresentation}) of the group $\Gamma$ as a finite set of generators and relations expressed as words in the generators, and returns all possible normal subgroups $\Gamma_\PBC$ of index $N$ up to a specified finite maximal index $N_\text{max}$. The output for each normal subgroup is in the form of a finite generating set $W$ whose elements are expressed as words in the generators $\gamma_1,\ldots,\gamma_4$ of $\Gamma$, such that $\Gamma_\PBC=\langle W\rangle$. This is isomorphic to the free group on $W$ modulo the set of relations in $W$ that descend from the (unique) relation in $\Gamma$. In practice, this latter relation is automatically satisfied when working with the $PSU(1,1)$ representation of the generators, thus the action of $\Gamma_\PBC$ on $\mathbb{H}$ is simply obtained by repeated application of the words in $W$.

The fact that $\Gamma_\PBC$ is a subgroup of index $N$ in $\Gamma$ implies the (right) coset decomposition
\begin{align}\label{CosetDecomp}
\Gamma=\Gamma_\text{PBC}\sqcup\Gamma_\text{PBC}g_2\sqcup\cdots\sqcup\Gamma_\text{PBC}g_N,
\end{align}
where $\sqcup$ denotes disjoint union, and the set
\begin{align}
T=\{g_1=e,g_2,\ldots,g_N\}\subset\Gamma
\end{align}
of coset representatives, where $e$ designates the identity element, is called a (right) {\it transversal} for $\Gamma_\PBC$ in $\Gamma$. (For a normal subgroup, right and left cosets are equivalent, and we will hereafter omit this distinction.) The fact that $\Gamma$ tiles all of $\mathbb{H}$ with copies of $\c{D}$ can be expressed as $\mathbb{H}=\sqcup_{\gamma\in\Gamma}\gamma\c{D}$. Likewise, Eq.~(\ref{CosetDecomp}) implies that
\begin{align}
\mathbb{H}=\bigsqcup_{\gamma_\PBC\in\Gamma_\PBC}\gamma_\PBC\c{C},
\end{align}
i.e., $\Gamma_\PBC$ tiles all of $\mathbb{H}$ with copies of the cluster $\c{C}$, where
\begin{align}\label{C}
\c{C}=\bigsqcup_{i=1}^Ng_i\c{D}.
\end{align}
While the choice of transversal is not unique, since any $g_i\in T$ can always be left-multiplied by an arbitrary element of $\Gamma_\PBC$, a physical choice of transversal is one in which $\c{C}$ forms a connected region in $\mathbb{H}$. Indeed, for such a choice the resulting finite portion $\{z_i\equiv g_i(0),i=1,\ldots,N\}$ of hyperbolic lattice will form a connected graph once nearest-neighbor hopping is introduced. We will henceforth refer to clusters associated with such a choice of transversal as {\it connected clusters}, and will exclusively consider those; for technical reasons we also limit our study to connected clusters for which words of length 3 and up do not appear in the transversal ({\it SI Appendix}, Sec.~S1).

\begin{figure}[t]
\includegraphics[width=\columnwidth]{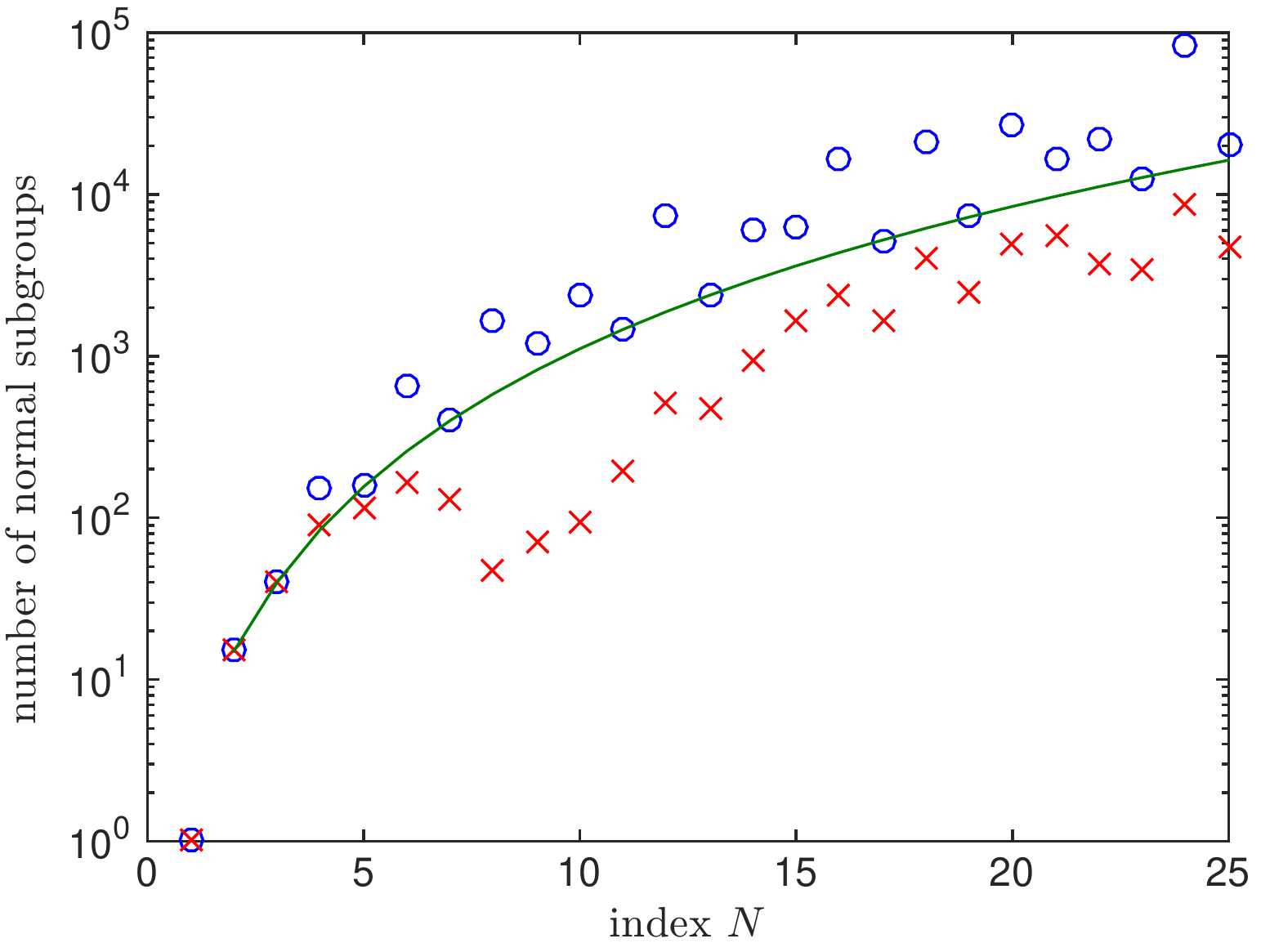}
\caption{Number of distinct normal subgroups $\Gamma_\PBC$ in $\Gamma$ as a function of index $N$, as computed by LINS. Blue circles: all normal subgroups; green line: plot of Eq.~(\ref{NSGp}); red crosses: normal subgroups giving rise to connected PBC clusters for which words of length 3 and longer are excluded from the transversal $T$.}
\label{fig:LINS}
\end{figure}

The number of normal subgroups grows rapidly with index, albeit nonmonotonically (Fig.~\ref{fig:LINS}, blue circles), and the computational time required to enumerate them grows as well. The number NSG$_p$ of normal subgroups of {\it prime} index $p$ can be determined analytically ({\it SI Appendix}, Sec.~S2):
\begin{align}\label{NSGp}
\text{NSG}_p=1+p+p^2+p^3,
\end{align}
which agrees with our computational results. We have performed computations using LINS up to $N_\text{max}=25$, which takes approximately one week on a single-CPU machine. For a given index, the number of normal subgroups giving rise to connected clusters of the type discussed above is roughly an order magnitude less than the total number of subgroups (Fig.~\ref{fig:LINS}, red crosses). However, the number of connected clusters also grows rapidly, and although the growth is again nonmonotonic, we hypothesize connected PBC clusters can be found for arbitrarily large $N$.

\begin{figure}[t]
\includegraphics[width=\columnwidth]{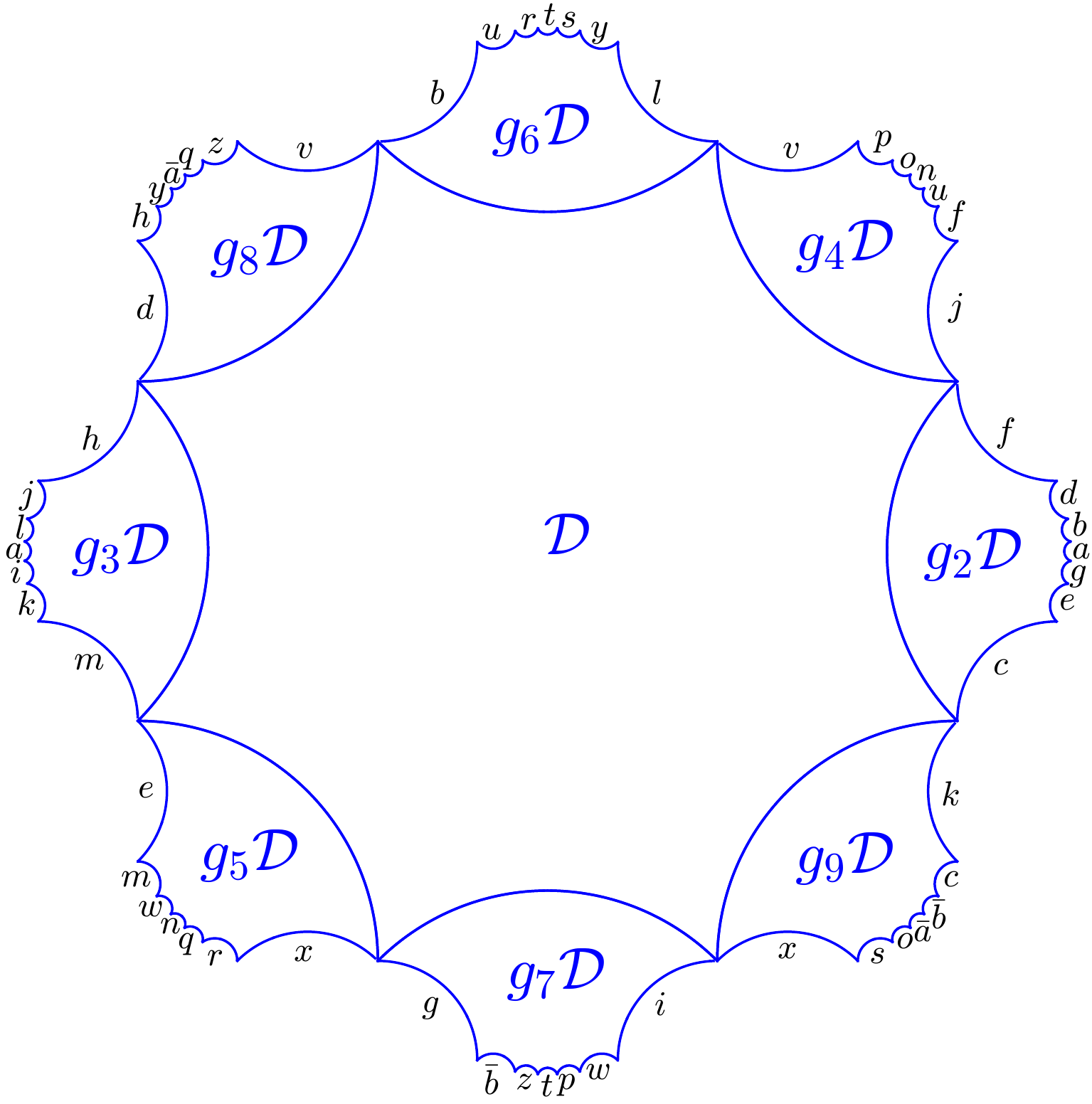}
\caption{Example of connected cluster of size $N=9$ and its pairwise edge identifications implied by $\Gamma_\PBC$.}
\label{fig:n9}
\end{figure}

We plot in Fig.~\ref{fig:n9} an example of connected cluster with $N=9$ unit cells. While all $N=9$ clusters consist of the disjoint union of the central octagon $\c{D}$ and its eight nearest neighbors, and are thus identical as subsets of the Poincar\'e disk, they differ in their pairwise edge identifications. The latter depend on the particular index-9 subgroup $\Gamma_\PBC$ considered and can be reconstructed from it. In {\it The hopping matrix} below, we derive the fact that two octagons $g_i\c{D}$ and $g_j\c{D}$ are nearest neighbors on a PBC cluster if there exists a group element $\gamma_\PBC\in\Gamma_\PBC$ such that $g_i\gamma_\alpha=\gamma_\PBC g_j$, where $\gamma_\alpha\in\{\gamma_1,\ldots,\gamma_4,\gamma_1^{-1},\ldots,\gamma_4^{-1}\}$. To determine if two octagons $g_i\c{D}$ and $g_j\c{D}$ share a common boundary in the compactified surface, one can then form all eight group elements $g_i\gamma_\alpha g_j^{-1}$ and check if they belong to $\Gamma_\PBC$, which is easily done in GAP. Excluding the common boundary that would persist in the presence of open boundary conditions, one can systematically determine the 28 orientation-preserving pairwise identifications that turn the 56-sided hyperbolic polygon in Fig.~\ref{fig:n9} into a genus-10 surface. This can be done for any PBC cluster.

\subsection*{The hopping matrix}
\label{sec:hopping}

Having developed an algorithm to systematically construct connected PBC clusters of (in principle) arbitrary size $N$, we turn to the construction of a tight-binding Hamiltonian on this cluster, which has the general form
\begin{align}\label{HPBC}
\c{H}_\text{PBC}=\sum_{i,j=1}^N H_{ij}c_i^\dag c_j,
\end{align}
in second quantization, where $c_i^{(\dag)}$ annihilates (creates) a particle on site $i$ of the cluster with coordinate $z_i=g_i(0)$, $g_i\in T$ in the Poincar\'e disk. As we are only interested in single-particle physics, the statistics of the creation/annihilation operators is irrelevant; our goal is only to construct and diagonalize the $N\times N$ hopping matrix $H_{ij}$. We focus here on nearest-neighbor hopping, where the relevant notion of distance is hyperbolic distance; extensions to longer-range hopping will be seen to be straightforward. We thus wish to set $H_{ij}=-1$ if $i$ and $j$ are nearest neighbors, and $H_{ij}=0$ otherwise. As discussed in Ref.~\cite{boettcher2021}, to find the 8 nearest neighbors of a site $z_i$ on the infinite $\{8,8\}$ lattice, one $\Gamma$-translates $z_i$ back to the origin $z=0$, applies any of the 8 length-1 words $\gamma_j,\gamma_j^{-1}$, $j=1,\ldots,4$ to $z=0$, and $\Gamma$-translates back. On an infinite lattice, the 8 nearest neighbors $z_{j_\alpha}$, $\alpha=1,\ldots,8$ of $z_i=g_i(0)$ are then
\begin{align}
z_{j_{\alpha}}=(g_i\gamma_\alpha g_i^{-1})g_i(0)=g_i\gamma_\alpha(0),
\end{align}
where $\gamma_\alpha\in\{\gamma_1,\ldots,\gamma_4,\gamma_1^{-1},\ldots,\gamma_4^{-1}\}$. Indeed, one can check using the explicit $PSU(1,1)$ matrices (\ref{alphabeta}) that the nearest-neighbor hyperbolic distance $\ell$ is\footnote{We have chosen our lattice sites to lie at the {\it centers} of hyperbolic octagons, but could have equally well chosen them to lie at the {\it vertices} of those octagons: the $\{8,8\}$ lattice is self-dual, and the nearest-neighbor vertex-vertex distance is again $\ell=\cosh^{-1}(5+4\sqrt{2})$.}:
\begin{align}
\ell=d(z_i,z_{j_\alpha})=\cosh^{-1}(5+4\sqrt{2})\approx 3.057,
\end{align}
and is the same for all $i$, given that $d(\gamma(z),\gamma(z'))=d(z,z')$ for any hyperbolic isometry $\gamma$.

While computation of the hyperbolic distance allows us to find the nearest neighbors of a site on an infinite lattice, this is not sufficient on a finite PBC cluster, since sites that appear further apart than $\ell$ in $\mathbb{H}$, e.g., on opposite edges of the cluster, may be nearest neighbors on the compactified surface $Y_N=\mathbb{H}/\Gamma_\PBC$. We thus need a notion of distance on the PBC cluster, i.e., distance modulo elements of $\Gamma_\PBC$, which can be formally defined as:
\begin{align}
d_\PBC(z_i,z_j)=\min_{\gamma_\PBC\in\Gamma_\PBC} d(z_i,\gamma_\PBC z_j).
\end{align}
Nearest neighbors on the cluster are then those pairs $z_i,z_j$ such that $d_\PBC(z_i,z_j)=\ell$.

To implement this distance function in practice, it is useful to think of the $N$ sites $z_i$ of the cluster as elements in the factor group $\Gamma/\Gamma_\PBC$, i.e., the cosets $\Gamma_\PBC g_i$. The sites $z_i=g_i(0)$ and $z_j=g_j(0)$ are nearest neighbors on the cluster if there exists a $\gamma_\PBC\in\Gamma_\PBC$ such that $g_i\gamma_\alpha(0)=\gamma_\PBC g_j(0)$. Left-multiplying by the group $\Gamma_\PBC$ on both sides of this equality, we obtain the requirement that
\begin{align}\label{Eq18}
\Gamma_\PBC g_i\gamma_\alpha(0)=\Gamma_\PBC g_j(0).
\end{align}
Since there is a unique hyperbolic transformation which connects two points in $\mathbb{H}$, and cosets form a disjoint partition of $\Gamma$ [Eq.~(\ref{CosetDecomp})], this implies the equality $\Gamma_\PBC g_i\gamma_\alpha=\Gamma_\PBC g_j$. In other words, sites $z_i$ and $z_j$ are nearest neighbors on the cluster if $g_j$ and $g_i\gamma_\alpha$ belong to the same coset of $\Gamma_\PBC$ in $\Gamma$. GAP routines for finitely presented groups~\cite{GAP4} allow one to determine whether two elements of such a group $\Gamma$ belong to the same right coset of a subgroup $\Gamma_\PBC$ of $\Gamma$. In practice, for each $g_i\in T$ and $\alpha=1,\ldots,8$, we utilize those routines to determine which element $g_j\in T$ belongs to the same coset as $g_i\gamma_\alpha$, and assign $H_{ij}=-1$. Since each generator and its inverse both appear in the set $\{\gamma_\alpha\}$, the resulting (real) hopping matrix is automatically symmetric, and thus defines a valid tight-binding Hamiltonian on the PBC cluster.

\subsection*{Fuchsian translation symmetry in finite size}
\label{sec:trans}

As reviewed earlier, for a 1D Euclidean chain of $N$ sites with PBC, the infinite translation group $G=\mathbb{Z}$ reduces to the finite group $G/G_\PBC=\mathbb{Z}/N\mathbb{Z}=\mathbb{Z}_N$, the cyclic group of order $N$. Ignoring symmetries other than translational (e.g., point-group symmetries), this is the symmetry group of the finite lattice. Viewed formally, a translation on the cluster acts as some permutation of the $N$ sites: there is a faithful (injective) homomorphism $\c{U}:G/G_\PBC\rightarrow\c{S}_N$, where $\c{S}_N$ is the permutation group on $N$ elements. In practice, to each element $g\in\{0,1,\ldots,N-1\}\cong\mathbb{Z}_N$, one can assign a permutation matrix $\c{U}(g)$ which acts by multiplication on a column vector of sites $x=(1,2,\ldots,N)^T$:
\begin{align}\label{1Dpermutation}
\c{U}(0)&=\mathbb{I}_N,\hspace{5mm}
\c{U}(1)=\left(\begin{array}{ccccc}
0 & 1 & 0 & \cdots & 0 \\
0 & 0 & 1 & \cdots & 0 \\
\vdots & \vdots & \vdots & \ddots & \vdots \\
0 & 0 & 0 & \cdots & 1 \\
1 & 0 & 0 & \cdots & 0
\end{array}\right),\nn\\
\c{U}(2)&=\c{U}(1)^2,\hspace{5mm}\ldots,\hspace{5mm}
\c{U}(N-1)=\c{U}(1)^{N-1},
\end{align}
where $\mathbb{I}_N$ denotes the $N\times N$ identity matrix. Translation symmetry on the PBC cluster is the statement that the hopping matrix commutes with the translation matrices $\c{U}(g)$ for all $g\in G/G_\PBC$. 

In the hyperbolic case, the factor group $\Gamma/\Gamma_\PBC$ is the group of residual $\Gamma$-translations on the PBC cluster. By Cayley's theorem, any finite group of order $N$ admits a faithful homomorphism to the permutation group $\c{S}_N$, thus we look for the $N\times N$ matrix representation of such a homomorphism:
\begin{align}\label{homomorphism}
\c{U}:\Gamma/\Gamma_\PBC\rightarrow\c{S}_N.
\end{align}

Before constructing the translation matrices $\c{U}$ in finite size, we first review the concept of Fuchsian translation symmetry on an infinite hyperbolic lattice~\cite{maciejko2020,boettcher2021}. For nearest-neighbor hopping on the $\{8,8\}$ lattice, the second-quantized Hamiltonian is:
\begin{align}\label{infiniteH}
\c{H}=\sum_{ij}H_{ij}c_i^\dag c_j^{\phantom{\dag}}=-\sum_{\gamma\in\Gamma}\sum_\alpha c_{\gamma(0)}^\dag c_{\gamma\gamma_\alpha(0)}^{\phantom{\dag}},
\end{align}
where $\gamma_\alpha\in\{\gamma_1,\ldots,\gamma_4,\gamma_1^{-1},\ldots,\gamma_4^{-1}\}$ as before. Translation symmetry on this infinite lattice is the statement that $[\c{T}_\gamma,\c{H}]=0$ for all $\gamma\in\Gamma$, where the translation operators $\c{T}_\gamma$ act on creation/annihilation operators as:
\begin{align}\label{TforGamma}
\c{T}_\gamma^{\phantom{}} c_z^{(\dag)}\c{T}_\gamma^{-1}=c_{\gamma(z)}^{(\dag)}.
\end{align}
Indeed, using the rearrangement lemma, it is easy to prove that $\c{T}_\gamma\c{H}\c{T}_\gamma^{-1}=\c{H}$.

On a finite PBC cluster, recall that the sites $z_i$ are best viewed as elements of $\Gamma/\Gamma_\PBC$, which are the $N$ cosets $\Gamma_\PBC g_i$ denoted for simplicity by $[g_i]$. Using this notation, the tight-binding Hamiltonian (\ref{HPBC}) can thus be written as
\begin{align}
\c{H}_\PBC=\sum_{i,j=1}^N H_{ij}c_{[g_i]}^\dag c_{[g_j]}^{\phantom{\dag}}=-\sum_{[g_i]\in\Gamma/\Gamma_\PBC}\sum_\alpha c_{[g_i]}^\dag c_{[g_i\gamma_\alpha]}^{\phantom{\dag}},
\end{align}
which parallels the structure of (\ref{infiniteH}), but where the finite hopping matrix is
\begin{align}\label{hopping2}
H_{ij}=-\sum_\alpha \delta_{[g_j],[g_i\gamma_\alpha]},
\end{align}
as described operationally in the paragraph following Eq.~(\ref{Eq18}). The action (\ref{TforGamma}) of the infinite group on the creation/annihilation operators is replaced by an action of $\Gamma/\Gamma_\PBC$,
\begin{align}
\c{T}_{[g_k]}^{\phantom{}}c_{[g_j]}^{(\dag)}\c{T}_{[g_k]}^{-1}=c_{[g_kg_j]}^{(\dag)}=\sum_i c_{[g_i]}^{(\dag)}\c{U}_{ij}([g_k]),
\end{align}
where
\begin{align}\label{Umatrix}
\c{U}_{ij}([g_k])=\delta_{[g_i],[g_kg_j]},
\end{align}
is the desired homomorphism (\ref{homomorphism}). In {\it SI Appendix}, Sec.~S3, we show that the matrices (\ref{Umatrix}) form a faithful representation of $\Gamma/\Gamma_\PBC$ and commute with the hopping matrix $H$. In practice, we construct the translation matrices for a given PBC cluster using the same GAP routines as for the hopping matrix: for each pair $j,k=1,\ldots,N$, we determine which element $g_i$ of the transversal belongs to the same coset as $g_kg_j$, and assign $\c{U}_{ij}([g_k])=1$.

\section*{Abelian clusters}
\label{sec:abelian}

In the 1D Euclidean example discussed earlier, the permutation matrices (\ref{1Dpermutation}) are in fact circulant matrices, which all mutually commute since they are all given by some positive power of the same matrix $\c{U}(1)$. Together with the hopping matrix, they form a mutually commuting set and can thus be simultaneously diagonalized. Since $\c{U}(g)^N=\mathbb{I}_N$ for all $g\in\{0,1,\ldots,N-1\}\cong\mathbb{Z}_N$, the $N$ eigenvalues $\chi^{(\lambda)}(g)$, $\lambda=0,\ldots,N-1$ of $\c{U}(g)$ are $N$th roots of unity. Explicitly, we have $\chi^{(\lambda)}(g)=e^{-i2\pi\lambda g/N}$, which is nothing but the Bloch phase factor associated with crystal momentum $k=2\pi\lambda/N$. In representation-theoretic terms, each $\lambda$ defines a 1D irrep $\chi^{(\lambda)}:\mathbb{Z}_N\rightarrow U(1)$. Since $\mathbb{Z}_N$ is abelian, this exhausts the set of all irreps. Finally, since the hopping matrix $H$ commutes with all $\c{U}(g)$, the eigenstates of $H$ are also eigenstates of $\c{U}(g)$, and thus obey Bloch's theorem: $\psi^{(\lambda)}(g^{-1}(x))=\chi^{(\lambda)}(g)\psi^{(\lambda)}(x)$.

\begin{figure}[t]
\includegraphics[width=\columnwidth]{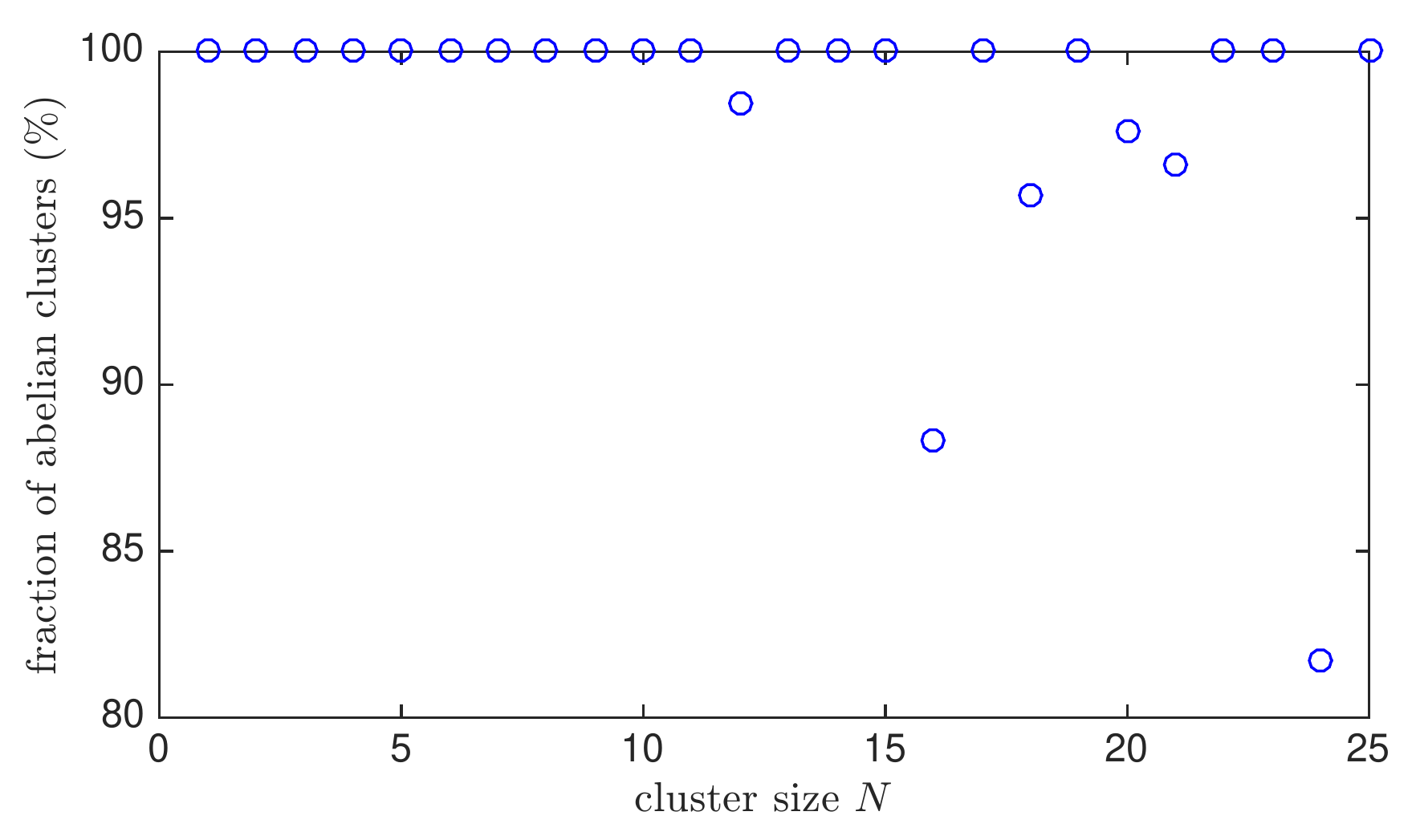}
\caption{Fraction of connected clusters for which the residual translation group $\Gamma/\Gamma_\PBC$ is abelian. For $N\leq N_\text{max}=25$, nonabelian clusters are found only at sizes $N=12,16,18,20,21,24$.}
\label{fig:abelian}
\end{figure}

In the hyperbolic case, the infinite translation group $\Gamma$ is nonabelian, thus we would not generally expect that the residual translation group $\Gamma/\Gamma_\PBC$ is abelian. According to this expectation, the translation matrices $\c{U}([g_k])$ would not mutually commute, and Hamiltonian eigenstates would not obey the $U(1)$ automorphic Bloch condition (\ref{AutomorphicBloch}). Surprisingly, for a large fraction of connected clusters, $\Gamma/\Gamma_\PBC$ {\it is in fact abelian} (Fig.~\ref{fig:abelian}). We hereafter refer to such clusters as {\it abelian clusters}\footnote{Bloch theory on abelian covers of a general class of connected graphs is also studied in Ref.~\cite{kollar2020}.}, and denote clusters for which $\Gamma/\Gamma_\PBC$ is nonabelian as {\it nonabelian clusters}. Out of the twenty-five distinct system sizes $N=1,\ldots,25$ we have investigated, only six admit nonabelian clusters, whereas all admit abelian clusters. Furthermore, for all system sizes studied, the proportion of abelian clusters is greater than 80\%. We also note that all clusters of prime size $p$ are necessarily abelian, since any finite group of prime order $p$ is isomorphic to $\mathbb{Z}_p$.

The second key result of the present work is thus that, despite Fuchsian translations being non-Euclidean in nature, PBC on finite hyperbolic lattices are possible such that the $U(1)$ automorphic Bloch condition proposed in Ref.~\cite{maciejko2020} {\it becomes exact and applies to all states in the spectrum}. More precisely, for abelian clusters, eigenstates of $H$ must obey the $U(1)$ automorphic Bloch theorem:
\begin{align}\label{U1Bloch}
\psi^{(\lambda)}(g_k^{-1}(z_i))=\chi^{(\lambda)}([g_k])\psi^{(\lambda)}(z_i),\hspace{5mm}[g_k]\in\Gamma/\Gamma_\PBC.
\end{align}
For such clusters, the $N$ translation matrices $\c{U}([g_k])$, $k=1,\ldots,N$ form a mutually commuting set and can be simultaneously diagonalized by some common transformation $P$:
\begin{align}\label{diagUabelian}
P\c{U}([g_k])P^{-1}=\left(\begin{array}{ccc}
\chi^{(1)}([g_k]) & & \\
& \ddots & \\
& & \chi^{(N)}([g_k])
\end{array}\right),
\end{align}
where each Bloch factor $\chi^{(\lambda)}([g_k])$, $\lambda=1,\ldots,N$ defines a 1D irrep $\chi^{(\lambda)}:\Gamma/\Gamma_\PBC\rightarrow U(1)$. Indeed, as in the Euclidean case, the translation matrices are $N\times N$ permutation matrices, thus their eigenvalues are roots of unity.

\subsection*{Discretization of the Jacobian}

In Ref.~\cite{maciejko2020}, we considered the automorphic Bloch condition $\psi(\gamma^{-1}(z))=\chi(\gamma)\psi(z)$ such that $\chi:\Gamma\rightarrow U(1)$ was a $U(1)$ irrep of the infinite group $\Gamma$. By the Narasimhan--Seshadri theorem~\cite{narasimhan1965} in rank 1, the space of all such irreps, also known as a character variety, forms a $2g$-dimensional torus $\Jac(\Sigma_g)\cong T^{2g}$, the Jacobian variety of the Riemann surface $\Sigma_g$. The space $\Jac(\Sigma_g)$ can be interpreted physically as the set of independent magnetic fluxes that can thread the $2g$ noncontractible cycles of the compactified unit cell of a $\{4g,4g\}$ lattice. For the Bolza lattice with $g=2$, there are four such fluxes, and we defined $\chi$ by its action on the generators of $\Gamma$: $\chi(\gamma_j)=\chi^*(\gamma_j^{-1})=e^{-ik_j}$, $j=1,\ldots,4$. Each component $k_j$ of the hyperbolic crystal momentum $\b{k}=(k_1,k_2,k_3,k_4)$ could then assume a continuous set of values in $[-\pi,\pi]/\sim$, with $\sim$ the antipodal map that identifies $\pm\pi$ in the interval.

For a finite PBC cluster, we expect by analogy with conventional band theory~\cite{SSP} that the crystal momentum becomes discretized. In the 1D Euclidean example above, this occurs because the irrep $\chi(g)=e^{-ikg}$, $k\in[-\pi,\pi]/\sim$ of $G=\mathbb{Z}$ is a valid irrep of $G/G_\PBC=\mathbb{Z}_N$ only if $k$ is an integer multiple of $2\pi/N$: otherwise, $\chi(g_\PBC)\neq 1$ for $g_\PBC\in G_\PBC=N\mathbb{Z}$. Likewise here, a $U(1)$ irrep $\chi$ of $\Gamma$ is only a valid irrep of $\Gamma/\Gamma_\PBC$ if $\chi(\gamma_\PBC)=1$ for $\gamma_\PBC\in\Gamma_\PBC$, which imposes a discretization condition on the hyperbolic crystal momentum $\b{k}$. Indeed, having obtained the $N$ eigenvalues $\chi^{(\lambda)}([g_k])$, $\lambda=1,\ldots,N$ for each $[g_k]$ by simultaneous diagonalization of the translation matrices $\c{U}$, the $N$ allowed values $\b{k}^{(\lambda)}=(k_1^{(\lambda)},k_2^{(\lambda)},k_3^{(\lambda)},k_4^{(\lambda)})$ of hyperbolic crystal momentum are obtained by considering elementary translations $[g_k]=[\gamma_j]$:
\begin{align}
\chi^{(\lambda)}([\gamma_j])=e^{-ik_j^{(\lambda)}},\hspace{5mm}j=1,\ldots,4,\hspace{5mm}\lambda=1,\ldots,N.
\end{align}
Since the $\chi^{(\lambda)}$ are roots of unity, the components $k_j^{(\lambda)}$ are necessarily rational multiples of $2\pi$, but the set of allowed points $\b{k}^{(\lambda)}$ in $\Jac(\Sigma_2)\cong T^4$ depends on the particular cluster considered, i.e., the particular normal subgroup $\Gamma_\PBC$. The Euclidean counterpart of this statement is a familiar fact encountered, e.g., in numerical exact diagonalization studies of quantum lattice Hamiltonians, where PBC clusters of different shapes allow for different discrete ``samplings'' of the Brillouin zone.

\begin{figure}[t]
\includegraphics[width=\columnwidth]{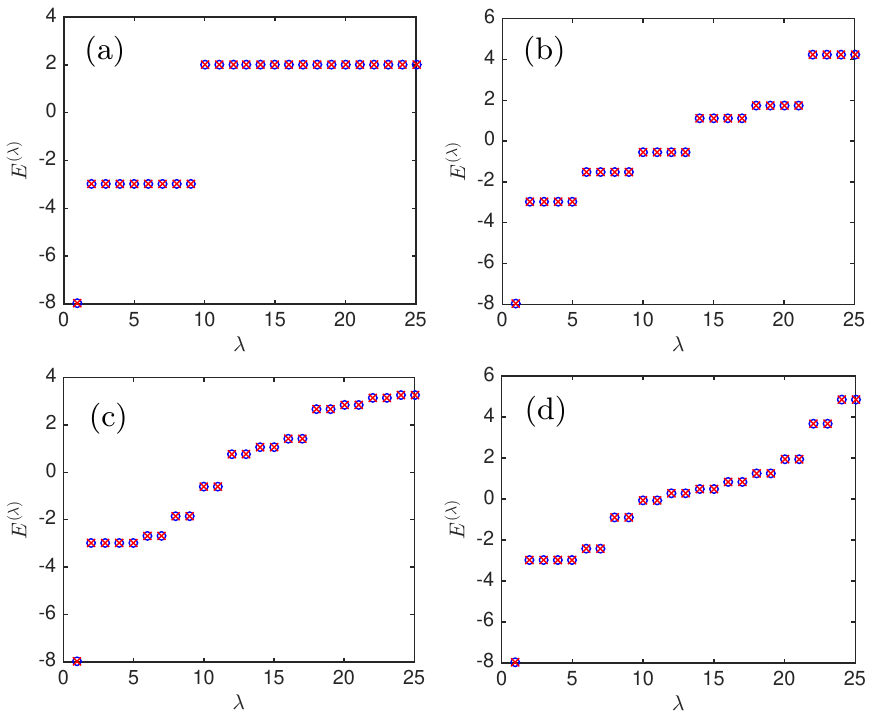}
 \caption{Comparison between the energy spectrum computed from exact diagonalization of the hopping matrix (red crosses) and from $U(1)$ hyperbolic band theory (blue circles), for four distinct $N=25$ clusters denoted (a), (b), (c), and (d).}
  \label{fig:n25spec}
\end{figure}

Substituting the $U(1)$ automorphic Bloch ansatz (\ref{U1Bloch}) in the Schr\"odinger equation $\sum_j H_{ij}\psi(z_j)=E\psi(z_i)$ for the nearest-neighbor Hamiltonian (\ref{hopping2}), we obtain the hyperbolic band structure in finite size,
\begin{align}\label{U1bandstructure}
E^{(\lambda)}&=E\left(\b{k}^{(\lambda)}\right)=-\sum_\alpha\chi^*([\gamma_\alpha])=-2\sum_{j=1}^4\cos k_j^{(\lambda)},\nn\\
&\lambda=1,\ldots,N.
\end{align}
In the last equality, we have used the fact that $\chi([\gamma_j^{-1}])=\chi([\gamma_j]^{-1})=\chi^*([\gamma_j])$. To assess the validity of our Bloch theorem (\ref{U1Bloch}), we can compare the band structure (\ref{U1bandstructure}) with the result of brute-force numerical diagonalization of the hopping matrix $H_{ij}$, not assuming any symmetries. We find that for all abelian clusters considered, there is an exact match between the energy spectra computed both ways (see Fig.~\ref{fig:n25spec} for an example for $N=25$). To get a sense of how well $\Jac(\Sigma_2)$ is sampled upon increasing the PBC cluster size $N$, we present in Sec.~S4 of {\it SI Appendix} density-of-states (DOS) histograms for three different cluster sizes. As the cluster size increases, they approximate the DOS of an infinite abelian cluster (see next section) increasingly well.

\subsection*{The commutator subgroup and the maximal abelian cover}

We have so far considered only the case where $\Gamma_\PBC$ is a normal subgroup of finite index, corresponding to a finite hyperbolic cluster. An example of normal subgroup of infinite index is the {\it commutator subgroup} $\Gamma^{(1)}=[\Gamma,\Gamma]$, which is the group freely generated by elements of the form $\gamma_i\gamma_j\gamma_i^{-1}\gamma_j^{-1}$, $i,j=1,\ldots,4$. Loosely speaking, the commutator subgroup measures the extent to which $\Gamma$ is nonabelian: the commutator subgroup of an abelian group is the trivial group with a single (identity) element. The commutator subgroup is also the smallest normal subgroup of $\Gamma$ such that the factor group is abelian; equivalently, the quotient $\Gamma/N$ with $N$ a normal subgroup of $\Gamma$ is abelian if and only if $\Gamma^{(1)}\subseteq N$. Thus for all abelian clusters encountered so far, one must have $\Gamma^{(1)}\subseteq\Gamma_\PBC$. Choosing $\Gamma_\PBC=\Gamma^{(1)}$ corresponds in fact to the compactification of an infinite subset of the original $\{8,8\}$ tessellation, and the space $Y_\infty=\mathbb{H}/\Gamma^{(1)}$ is the largest possible abelian cover of the Bolza surface $X=\mathbb{H}/\Gamma$. It is an abelian cover with infinitely many sheets, which we will call the {\it maximal abelian cover} of $X$. Geometrically, it is a Riemann surface of infinite genus.

The (infinite) group of residual translations on the maximal abelian cover $Y_\infty$ is the quotient $\Gamma/\Gamma^{(1)}$, known as the abelianization of $\Gamma$. By the Hurewicz theorem~\cite{AT}, $\Gamma/\Gamma^{(1)}$ is isomorphic to the first homology group $H_1(X,\mathbb{Z})$, which is abelian. For the $\{8,8\}$ lattice, we have $H_1(X,\mathbb{Z})\cong\mathbb{Z}^4$; more generally, for the $\{4g,4g\}$ lattice, we have $H_1(X,\mathbb{Z})\cong\mathbb{Z}^{2g}$. In physical terms, the maximal abelian cover is a subset of the original hyperbolic lattice that behaves as an infinite Euclidean lattice in $2g$ dimensions. As for finite abelian clusters, the $U(1)$ automorphic Bloch theorem holds exactly for the maximal abelian cover, but the hyperbolic momenta $\b{k}$ now form a continuous set mapping the entire Jacobian $\Jac(\Sigma_g)\cong T^{2g}$.

\section*{Nonabelian clusters: a nonabelian Bloch theorem}
\label{sec:nonabelian}

Having discussed abelian clusters, for which the $U(1)$ automorphic Bloch condition (\ref{AutomorphicBloch}) becomes a rigorous Bloch theorem (\ref{U1Bloch}), we next turn to nonabelian clusters, for which the residual translation group $\Gamma/\Gamma_\PBC$ is a nonabelian finite group of order $N$. One still obtains a homomorphism (\ref{homomorphism}), but the permutation matrices $\c{U}([g_k])$ do not mutually commute. However, they still commute with the hopping matrix $H$, thus we expect that eigenstates $\psi(z_i)$ of $H$ will form degenerate multiplets transforming according to irreps of $\Gamma/\Gamma_\PBC$:
\begin{align}\label{NABloch}
\psi^{(\lambda)}_\nu(g_k^{-1}(z_i))=\sum_{\mu=1}^{r_\lambda}\psi_\mu^{(\lambda)}(z_i)D_{\mu\nu}^{(\lambda)}([g_k]),\hspace{5mm}[g_k]\in\Gamma/\Gamma_\text{PBC}.
\end{align}
Here $\psi_\mu^{(\lambda)}$, $\mu=1,\ldots,r_\lambda$ are the $r_\lambda$ degenerate states belonging to irrep $\lambda$ of $\Gamma/\Gamma_\text{PBC}$, $r_\lambda$ is the dimension of that irrep, and $D^{(\lambda)}\in U(r_\lambda)$ are the unitary representation matrices. Equation (\ref{NABloch}) is the third key result of this work: namely that eigenstates of translationally invariant hopping Hamiltonians on finite hyperbolic lattices with PBC obey a nonabelian Bloch theorem. For 1D irreps such as the trivial representation, which is always present for any group, one has $r_\lambda=1$ and Eq.~(\ref{NABloch}) reduces to the abelian Bloch theorem (\ref{U1Bloch}), with $\chi^{(\lambda)}=D^{(\lambda)}$. For $\Gamma/\Gamma_\PBC$ nonabelian, there will also be irreps with $r_\lambda>1$, subject to the constraint that $\sum_{\lambda=1}^\c{N}r_\lambda^2=N$ where $\c{N}<N$ is the number of conjugacy classes of $\Gamma/\Gamma_\PBC$ (and thus also the number of irreps)~\cite{Tinkham}. For an abelian group, each element is in its own conjugacy class, thus $\c{N}=N$.

While the appearance of higher-dimensional irreps in the spectrum of $H$ is generally expected, one could contemplate the possibility that the multiplicity $a_\lambda$ of such an irrep $\lambda$, i.e., the number of times that a multiplet belonging to $\lambda$ appears in the spectrum, is in fact zero. However, we can easily show that {\it all} irreps must necessarily appear in the spectrum by recognizing that the translation matrices $\c{U}$ form what is known as the {\it regular representation} of $\Gamma/\Gamma_\PBC$. The regular representation of a group of order $N$ is the one derived from the defining representation of $\c{S}_N$ under the homomorphism of that group into $\c{S}_N$ implied by Cayley's theorem [recall Eq.~(\ref{homomorphism})]. The regular representation is reducible, and can be block-diagonalized by a suitable unitary transformation $P$,
\begin{align}\label{directsum}
P\c{U}([g_k])P^{-1}=\bigoplus_{\lambda=1}^\c{N}r_\lambda D^{(\lambda)}([g_k]),
\end{align}
i.e., decomposed into a direct sum of irreps $\lambda$, where the multiplicity is equal to the dimension $r_\lambda$ of irrep $\lambda$~\cite{Tinkham}. For an abelian group, we recover Eq.~(\ref{diagUabelian}): all irreps are 1D, and the regular representation $\c{U}$ can be fully diagonalized. By Schur's lemma, the matrix $PHP^{-1}$, which commutes with the $P\c{U}([g_k])P^{-1}$ by assumption, must necessarily be diagonal, with a number $r_\lambda$ of $r_\lambda$-fold degenerate energy eigenvalues $E_1^{(\lambda)},\ldots,E_{r_\lambda}^{(\lambda)}$. Since $\sum_{\lambda=1}^\c{N}r_\lambda^2=N$ from the general dimensionality theorem used earlier, this accounts for the entire spectrum. Thus provided irreps with $r_\lambda>1$ exist, as they do for $\Gamma/\Gamma_\PBC$ nonabelian, $r_\lambda$-fold degenerate multiplets obeying the nonabelian Bloch theorem (\ref{NABloch}) necessarily appear $r_\lambda$ times in the spectrum of nonabelian PBC clusters. For simplicity, we refer to eigenstates obeying the $U(1)$ automorphic Bloch theorem (\ref{U1Bloch}) as {\it abelian states}, and to eigenstates obeying the Bloch theorem (\ref{NABloch}) with $r_\lambda>1$ as {\it nonabelian states}. While abelian clusters only possess abelian states, nonabelian clusters possess states of both types. In {\it SI Appendix}, Sec.~S4, we show that nonabelian clusters can possess a significant fraction of abelian states.

\subsection*{Irrep decomposition of the finite-size spectrum: an explicit example}
\label{sec:NAexample}

\begin{figure}[t]
\centering\includegraphics[width=0.6\columnwidth]{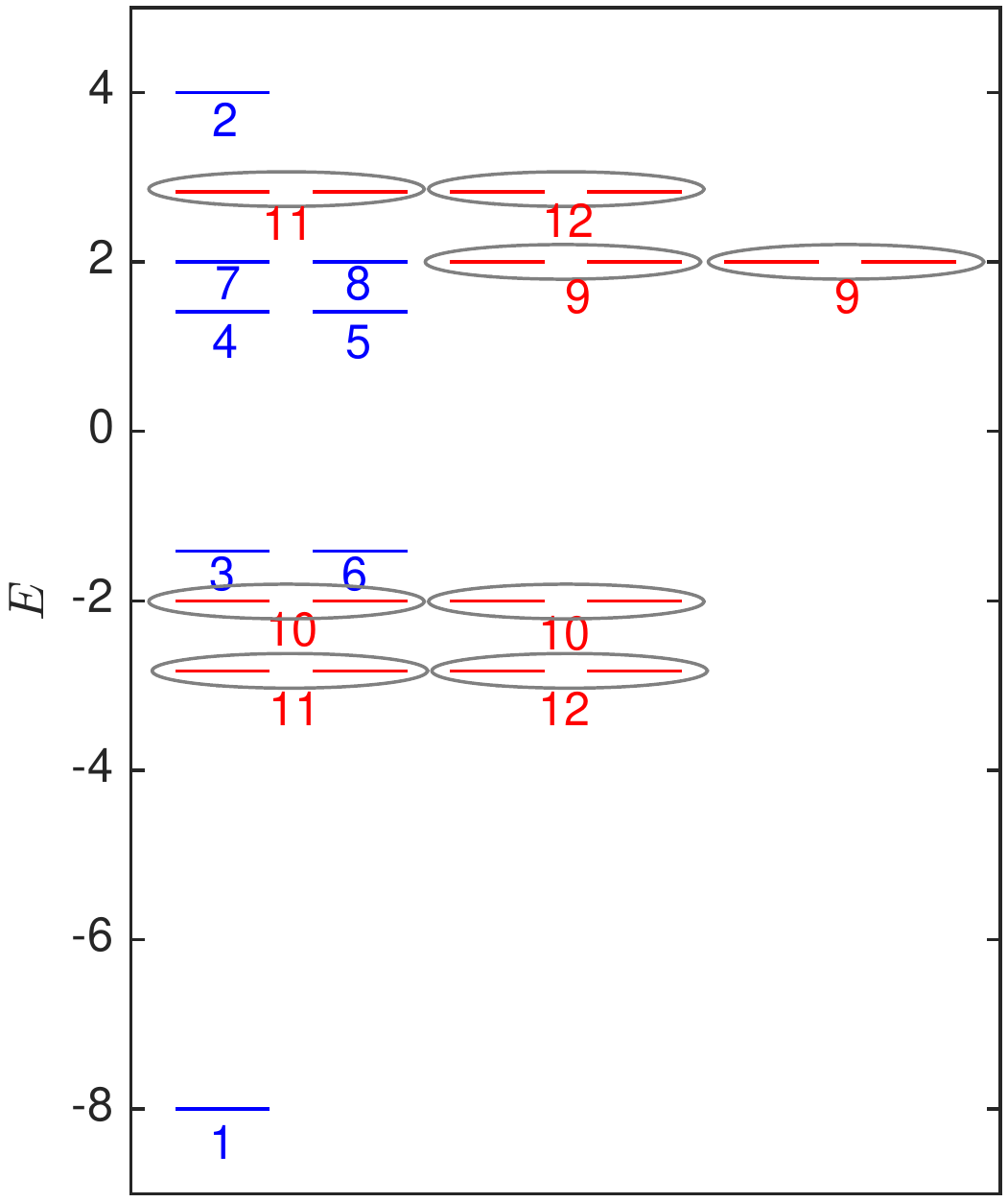}
\caption{Energy spectrum of the hopping Hamiltonian on the PBC cluster considered in {\it Irrep decomposition of the finite-size spectrum: an explicit example}. Blue: levels corresponding to abelian states; red: levels corresponding to nonabelian states. Numbers correspond to the irrep label $\lambda$, and grey ellipses denote eigenstates belonging to a (2D) nonabelian irrep.}
\label{fig:NAspec}
\end{figure}

Beyond computing the relative fraction of abelian vs nonabelian states in the spectrum of a given cluster, we now show with an explicit example how the finite-size spectrum may be fully characterized in terms of the irreps of $\Gamma/\Gamma_\PBC$; detailed calculations can be found in {\it SI Appendix}, Sec.~S5. We choose a specific nonabelian cluster of size $N=24$, and explicitly calculate in GAP the (irreducible) character table of its associated translation group $\Gamma/\Gamma_\PBC$. The group is found to have eight 1D irreps $\lambda=1,\ldots,8$ and four 2D irreps $\lambda=9,\ldots,12$. Accordingly, based on Fuchsian translation symmetry alone, we expect the spectrum of $H$ to consist of eight nondegenerate levels and eight two-fold degenerate multiplets (two copies of each 2D irrep), for a total of sixteen distinct eigenenergies (but $8+8\times 2=24$ eigenstates).

Numerically diagonalizing the hopping Hamiltonian $H$, we find only eight distinct eigenenergies, and observe four-fold and even six-fold degeneracies (Fig.~\ref{fig:NAspec}). This implies the presence of additional degeneracies beyond those required by Fuchsian translation symmetry, either accidental or arising from point-group symmetries~\cite{maciejko2020}, which are not considered here. To determine which of these observed degeneracies arise from Fuchsian translation symmetry, we construct projector matrices $\Pi^{(\lambda)}$ which project an arbitrary state onto irrep $\lambda$. We can then precisely determine to which irrep each of the twenty-four eigenstates of $H$ belongs (Fig.~\ref{fig:NAspec}, with 1D irreps in blue and 2D irreps in red). As a result, we can distinguish between degeneracies that are a consequence of the nonabelian Bloch theorem (\ref{NABloch}), and degeneracies that we will refer to as accidental (with the above caveat regarding point-group symmetries). The 1 and 2 irreps appear as nondegenerate levels, according to the generic expectation. The 3 and 6, 4 and 5, and 7 and 8 irreps appear in pairs, which is an accidental degeneracy. All those 1D irreps appear only once, as expected from our earlier discussion. The 2D 9, 10, 11, and 12 irreps each appear twice. The two copies of the 11 and 12 irreps appear at different energies, which is the generic scenario, but the two copies of the 9 and 10 irreps appear at the same energy, which is again an accidental degeneracy. Other accidental degeneracies are found between the $\lambda=7,8$ abelian states and the $\lambda=9$ nonabelian multiplets, and between the $\lambda=11$ and $\lambda=12$ nonabelian multiplets.

\subsection*{Discretization of the higher-rank moduli spaces}

The presence of nonabelian Brillouin zones in our nonabelian Bloch theorem manifests itself in terms of algebraic geometry through the full power of the Narasimhan--Seshadri theorem~\cite{narasimhan1965}. We now directly generalize unitary irreps $\chi:\Gamma\to U(1)$ in the automorphic Bloch condition to those of the form $\chi:\Gamma\to U(r)$, thereby producing a higher-rank character variety.  When $r>1$, the space of such irreps $\chi$ taken up to isomorphism does not admit the structure of a compact torus.  Rather, the moduli space is a $(2r^2(g-1)+2)$-dimensional manifold $\c{N}(\Sigma_g,U(r))$ that lacks the compactness and the group structure inherent to the rank-$1$ character variety, namely $\c{N}(\Sigma_g,U(1))\cong\Jac(\Sigma_g)$. Various aspects of its topology and geometry are well known (e.g.~\cite{thaddeus1995}). The stability condition on $\c{N}(\Sigma_g,U(r))$ was first introduced by Mumford in Ref.~\cite{mumford1963}; stability conditions have become familiar tools in algebraic geometry that permit geometers to construct topologically-nice spaces from quotients by equivalence relations and group actions.

The lack of compactness of $\c{N}(\Sigma_g,U(r))$ is corrected in a very mild way by admitting reducible representations.  However, the overall structure is still not toroidal. Indeed, $\c{N}(\Sigma_g,U(r))$ is the quotient of a Euclidean space by a lattice only when $r=1$. Now, by applying the classical Riemann--Hilbert correspondence (e.g. Ref.~\cite{mebkhout1980}), we can view $\c{N}(\Sigma_g,U(r))$ as a moduli space of reducible, flat $U(r)$ connections on $\Sigma_g$. The latter has previously appeared in physics in the semiclassical quantization of 2D Yang--Mills theory on Riemann surfaces~\cite{atiyah1983,witten1991} and in the canonical quantization of 3D Chern--Simons theory on a spacetime of the form $\Sigma_g\times\mathbb{R}$~\cite{witten1989,elitzur1989}. In turn, Narasimhan--Seshadri recasts this moduli space of flat connections as the moduli space $\c{M}(\Sigma_g,U(r))$ of semistable holomorphic vector bundles of (complex) rank $r$ and vanishing first Chern class, which enjoys its own connections to completely integrable Hamiltonian systems and quantization (e.g.~\cite{goldman1986,jeffrey1992}). Here, a holomorphic vector bundle $V$ on $\Sigma_g$ is said to be \emph{semistable} if each and every nonzero, proper subbundle $U\subsetneq V$ satisfies the following inequality: $c_1(U)/\mbox{rk}(U) \leq c_1(V)/r$, where $\mbox{rk}$ and $c_1$ denote the rank and the first Chern class (as an integer), respectively.  In other words, the \emph{normalized} first Chern class of each subbundle must not exceed that of the whole bundle.  This condition limits, in particular, the automorphisms available to a bundle, which is necessary for forming a topologically well-behaved moduli space. However, the bundles $V$ admitting at least one subbundle $U$ with $c_1(U)/\mbox{rk}(U) \leq c_1(V)/r$ are simultaneously the compactifying points of the topology as well as the (possibly) singular points.

It is worth noting that the correspondence $\c{N}(\Sigma_g,U(r))\cong\c{M}(\Sigma_g,U(r))$ is a diffeomorphism but not a complex-analytic isomorphism in general.  In other words, the correspondence presents two different complex manifold structures on the same differentiable manifold.  While $\c{N}(\Sigma_g,U(r))$ and $\c{M}(\Sigma_g,U(r))$ are topologically equivalent and equally suitable for capturing eigenstates within the hyperbolic band theory, and while $\c{N}(\Sigma_g,U(r))$ is physically appealing as a space of connections, $\c{M}(\Sigma_g,U(r))$ has a more rigid geometric structure that submits to tools from algebraic geometry that are not normally available for $\c{N}(\Sigma_g,U(r))$.  It is also important to note that $\c{N}(\Sigma_g,U(r))$ only depends on the topological information of the surface $\Sigma_g$, while the geometry of $\c{M}(\Sigma_g,U(r))$ depends on the Riemann surface structure on $\Sigma_g$\footnote{For further details concerning the Riemann--Hilbert correspondence, the Narasimhan--Seshadri correspondence, stability, differentiable versus complex structures on moduli spaces, and the larger nonabelian Hodge theory correspondence that these ideas fit into, we refer the reader to the survey by \cite{garciaraboso2015} for instance.}.

\begin{figure}[t]
\centering\includegraphics[width=0.5\columnwidth]{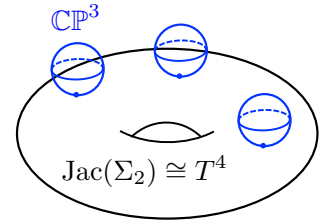}
 \caption{Schematic representation of the moduli space $\c{M}(\Sigma_2,U(2))$.}
  \label{fig:moduli}
\end{figure}

We now turn to a concrete example, the moduli space $\c{M}(\Sigma_2,U(2))$, that demonstrates the departure from the toroidal geometry of the abelian Brillouin zone and applies to the earlier discussion of 2D irreps for the Bolza surface. This moduli space is isomorphic to a bundle of copies of the complex projective space $\mathbb C\mathbb P^3$ over the Jacobian~\cite{narasimhan1969}.  In other words, the genus-$2$, $U(2)$ moduli space has a $U(1)$ direction that is the Jacobian and an $SU(2)$ direction whose geometry is more akin to that of the sphere (Fig.~\ref{fig:moduli})---indeed, the simplest complex projective space, $\mathbb C\mathbb P^1$, is exactly the Riemann sphere.  These $\mathbb C\mathbb P^3$ fibres are positively curved, unlike a torus which is geometrically flat.  For $\Jac(\Sigma_g)=\c M(\Sigma_g,U(1))$, the $SU(1)$ factor is trivial and thus we only detect the toroidal geometry of the Jacobian. In {\it SI Appendix}, Sec.~S6, we provide further information on the geometry of the higher-rank moduli spaces.

Now, let us denote the normal subgroup $\Gamma_\PBC\triangleleft\Gamma$ as $\Gamma'_N$ to emphasize its finite index $N$, where $\Gamma=\pi_1(\Sigma_g)$ as before.  Let $G_N=\Gamma/\Gamma'_N$, noting that $|G_N|=N$. We may construct a new Riemann surface $Y_N = \mathbb H / \Gamma'_N$, which projects onto $\Sigma_g$ as an $N$-fold Galois cover.  We use $f_N$ to denote the covering map $Y_N\to X$, where we use $X=\Sigma_g$ for simplicity.  (As discussed previously, the genus of $Y_N$ depends on $g$ and $N$ in a predictable way, as per the Riemann--Hurwitz theorem.)  Observe that we can recover $\Sigma_g$ as $Y_N/G_N$, where $G_N$ has the interpretation as a group of deck transformations for the cover.  As $N$ increases, the surfaces $Y_N$ and the groups $G_N$ can be regarded as a sequence of approximations to $\mathbb H$ and $\Gamma$, respectively.  On the moduli spaces side, we are replacing the character variety $\mbox{Irrep}(\Gamma,U(r))/U(r)$ with $\mbox{Irrep}(G_N,U(r))/U(r)$, which represents a discretization of the character variety.  This is an example of how using Narasimhan--Seshadri is helpful.  By viewing this in terms of moduli spaces of holomorphic bundles, we note that a rank-$r$ bundle $V$ on $\Sigma_g$ has a pullback $f_N^*V$ to $Y_N$.  The new bundle $f_N^*V$ has the same rank as $V$ and, after tensoring by a line bundle, has vanishing first Chern class.  (We will use $f_N^*V$ for this twisted bundle without ambiguity.)  Furthermore, stability is preserved under pulling back, and so $f_N^*V$ belongs to $\c{M}(Y_N,U(r))$, and thus a stable bundle on $X$ induces one on $Y_N$.  Now, if $V$ came from $\mbox{Irrep}(G_N,U(r))/U(r)$ specifically, then $V$ arises from a representation of $\Gamma$ that sends $\Gamma'_N$ to the identity in $U(r)$.  This means that the bundle $f_N^*V$ must be trivial except for possibly around the branch points of $f_N:Y_N\to X$, and so the moduli problem reduces to one on a divisor (a finite set of points) in $Y_N$.  In other words, $\mbox{Irrep}(G_N,U(r))/U(r)$ is discrete and we may thus perform, for arbitrary rank, similar explicit band-theoretic calculations as done in ranks $1$ and $2$ in the previous section.  We leave such systematic calculations for future exploration.

\section*{Summary and outlook}
\label{sec:conclusion}

In summary, we have extended the hyperbolic band theory of Ref.~\cite{maciejko2020} in several significant ways. First, based on earlier work of Sausset and Tarjus, we have generalized the notion of PBC for finite lattices from the Euclidean to the hyperbolic context, and shown that such a notion is compatible with the automorphic Bloch condition proposed in Ref.~\cite{maciejko2020}. In both the Euclidean and hyperbolic contexts, a finite PBC cluster with $N$ sites corresponds to a choice of normal subgroup $\Gamma_\PBC$ of finite index $N$ in the translation group $\Gamma$. We have used a mathematical algorithm, the low-index normal subgroups procedure, to systematically enumerate all possible PBC clusters of the $\{8,8\}$ lattice up to $N=25$. We then showed that the group of residual translations on the cluster is the factor group $\Gamma/\Gamma_\PBC$, a finite group of order $N$, and constructed nearest-neighbor hopping Hamiltonians invariant under this group. Although we have focused on the $\{8,8\}$ lattice, our constructions are straightforwardly generalized to all hyperbolic lattices for which a strictly hyperbolic, co-compact translation group $\Gamma$ associated with its underlying hyperbolic Bravais lattice~\cite{boettcher2021} can be identified.

Second, we established that for the majority of PBC clusters, $\Gamma/\Gamma_\PBC$ is in fact abelian, and the automorphic Bloch ansatz of Ref.~\cite{maciejko2020} with $U(1)$ factors of automorphy becomes exact. As with Euclidean lattices, the hyperbolic crystal momentum $\b{k}\in\Jac(\Sigma_g)\cong T^{2g}$ becomes discrete in finite size, with components valued in $2\pi\mathbb{Q}$. There exists in fact an infinite PBC cluster, corresponding to $\Gamma_\PBC$ equal to the commutator subgroup of $\Gamma$, which behaves as a Euclidean lattice in $2g$ dimensions. For this particular infinite cluster, $U(1)$ hyperbolic band theory is again exact, but this time with a continuous crystal momentum.

Third, we showed that for certain PBC clusters considered, $\Gamma/\Gamma_\PBC$ is nonabelian, and $U(1)$ factors of automorphy are not sufficient to describe the entire spectrum. Rather, we showed that some eigenstates obey a nonabelian Bloch theorem: they belong to degenerate multiplets, and transform into each other under Fuchsian translations. The analog of the discretization of crystal momentum in this case is the selection of discrete points from an otherwise continuous space, the moduli space $\c{M}(\Sigma_g,U(r))$ of stable holomorphic vector bundles of rank $r$, with $r>1$. This classic object in modern algebraic geometry, isomorphic to the space of inequivalent $U(r)$ irreps of $\Gamma$, emerges naturally from our construction in the infinite-size limit, and generalizes the Jacobian torus that only parametrizes $U(1)$ representations.

\begin{figure}[t]
\includegraphics[width=\columnwidth]{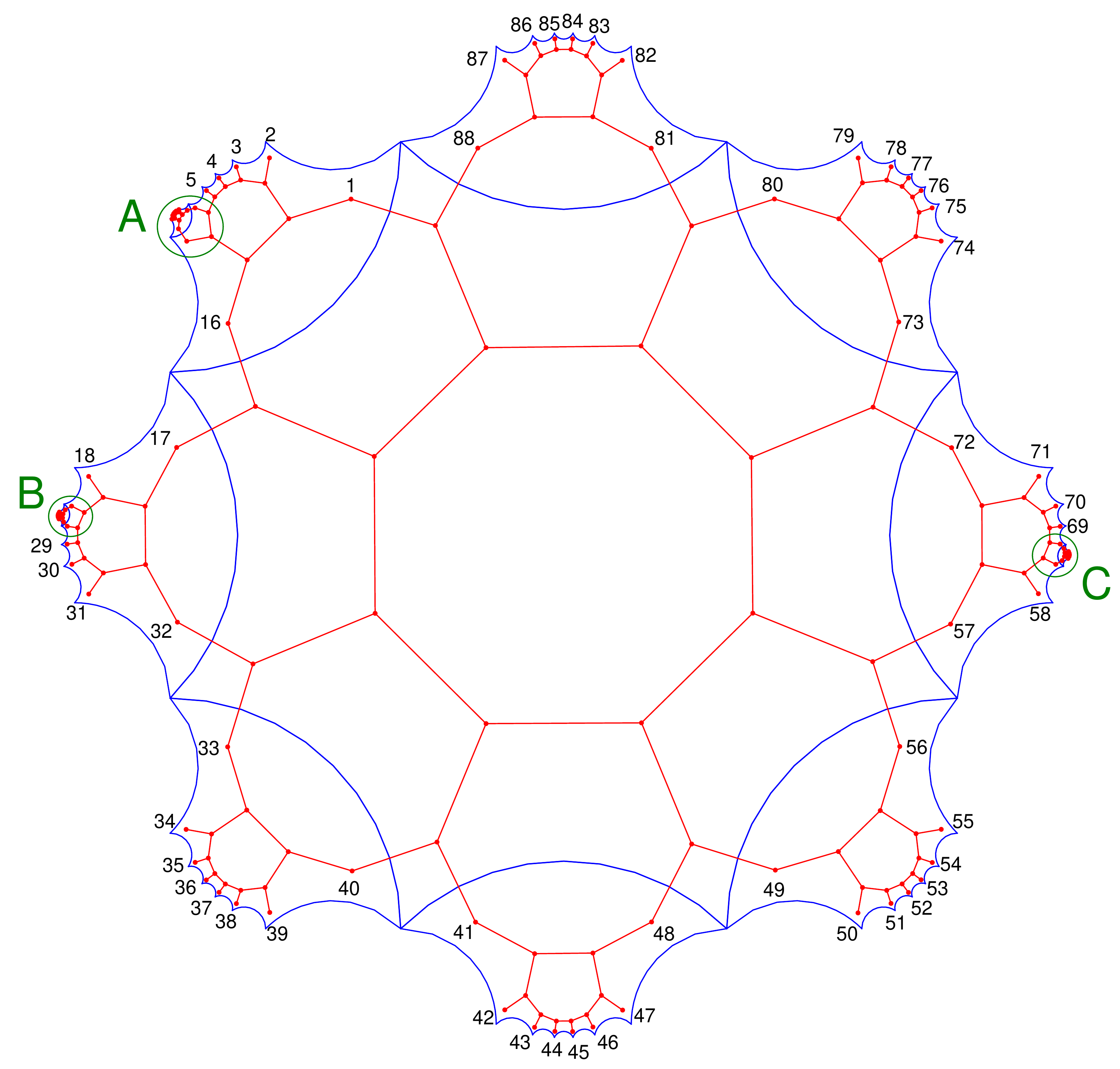}
 \caption{Proposed realization of abelian/nonabelian Bloch states using CQED or electric circuit implementations. A cluster of 192 sites (red vertices) of the $\{8,3\}$ lattice corresponds to $N=12$ Bolza unit cells (blue octagons) with 16 sites each. The 88 boundary sites can be wired together in two different ways to achieve either abelian or nonabelian PBC; see {\it SI Appendix}, Sec.~S7 for the list of boundary identifications and close-ups of the A, B, C boundary octagons (green).}
  \label{fig:CQED}
\end{figure}

As a concrete experimental proposal to realize the novel physics discussed in this work, we propose fabricating a device based on the $\{8,3\}$ lattice (Fig.~\ref{fig:CQED}); the heptagonal version of this tiling was used as layout graph in the CQED experiments of Ref.~\cite{kollar2019}. The kagome-like line graphs of such lattices, which the latter technology effectively implements, are of intrinsic interest due to their unusual flat bands~\cite{kollar2019,kollar2019b}. The proposed device contains $N=12$ unit cells of the Bolza lattice, the smallest value of $N$ at which nonabelian Bloch states appear (see Fig.~\ref{fig:abelian}). There are 16 sites per unit cell, for a total of 192 sites. In the CQED implementation, each site is decorated with a capacitive coupler and three coplanar waveguide resonators; with electric circuits, each site is a distinct node. While only lattices with open (Dirichlet) boundary conditions were considered in previous experiments~\cite{kollar2019,lenggenhager2021}, here we wish to exploit the flexibility of the CQED/electric circuit platforms to engineer PBC.  In {\it SI Appendix}, Sec.~S7, we give detailed prescriptions for connecting the 88 boundary sites in two different ways, yielding abelian and nonabelian PBC, respectively.

Finally, we indicate possible avenues for future theoretical research. First, finding an explicit parametrization of the irreducible $U(r)$ representation matrices in the nonabelian Bloch theorem (\ref{NABloch}) is an important question for future research. The genus-2, rank-2 case is a promising starting point, since the geometry of the associated moduli space is in principle known. Second, it would be interesting to explore the effect of threading global fluxes through the $2h$ cycles of the compactified PBC cluster, i.e., considering twisted PBC. For 2D Euclidean lattices, the space of such fluxes is $\Jac(T^2)\cong T^2$ regardless of system size. For a square lattice, inserting a pair of global fluxes $(\phi_x,\phi_y)$ in the $x$ and $y$ directions, respectively, leads to a shift of the crystal momentum $(k_x,k_y)\rightarrow(k_x+\phi_x,k_y+\phi_y)$. By contrast, for hyperbolic lattices, the space of global $U(1)$ fluxes is $\Jac(\Sigma_h)\cong T^{2h}$, whose dimension grows with the size of the system (recall Eq.~(\ref{RH}) and the surrounding discussion). Precisely how the quantized hyperbolic crystal momenta $\b{k}\in\Jac(\Sigma_g)$ are shifted upon tuning such global fluxes, and whether nonabelian global fluxes can also be inserted, are interesting questions for further research.   From the algebro-geometric point of view, the intricate (co)homology of nonabelian global fluxes in $\c{M}(\Sigma_g,U(r))$ may carry physical meaning worthy of investigation in this context.  At the same time, the existence of non-smooth values of the crystal momentum, corresponding to semistable (but not stable) vector bundles in $\c{M}(\Sigma_g,U(r))$, is a compelling new feature of the hyperbolic band theory that should be understood, as should the role of turning on nonzero values of the first Chern class of a stable bundle.  Lastly, the fact that stability for vector bundles is intimately tied to the Yang-Mills equations on a surface~\cite{atiyah1983} is suggestive of intriguing new connections between high-energy physics and condensed matter.

\acknow{We acknowledge useful discussions with I. Boettcher, A. Chen, M. Conder, A. Gorshkov, D. Holt, A. Hulpke, A. Koll\'ar, F. Marsiglio, R. Mazzeo, J. Szmigielski, R. Thomale, A. Topaz, and L. K. Upreti. While working on the research reported in this paper, J.M. was supported by NSERC Discovery Grants \#RGPIN-2020-06999 and \#RGPAS-2020-00064; the Canada Research Chair (CRC) Program; CIFAR; the Government of Alberta's Major Innovation Fund (MIF); and the University of Alberta. S.R. was supported by NSERC Discovery Grant \#RGPIN-2017-04520; the Canada Foundation for Innovation John R. Evans Leaders Fund; the GEAR Network (National Science Foundation grants DMS 1107452, 1107263, 1107367 \emph{RNMS: Geometric Structures and Representation Varieties}); and the University of Saskatchewan.  Both J.M. and S.R. were supported by the Tri-Agency New Frontiers in Research Fund (NFRF, Exploration Stream) and the Pacific Institute for the Mathematical Sciences (PIMS) Collaborative Research Group program.}

\showacknow{} 

\bibliography{PBC}

\end{document}


\newcommand{\tr}{\mathop{\mathrm{tr}}}
\newcommand{\bsigma}{\boldsymbol{\sigma}}
\renewcommand{\Re}{\mathop{\mathrm{Re}}}
\renewcommand{\Im}{\mathop{\mathrm{Im}}}
\renewcommand{\b}[1]{{\boldsymbol{#1}}}
\newcommand{\diag}{\mathrm{diag}}
\newcommand{\sign}{\mathrm{sign}}
\newcommand{\sgn}{\mathop{\mathrm{sgn}}}
\renewcommand{\c}[1]{\mathcal{#1}}
\newcommand{\Jac}{\mathop{\mathrm{Jac}}}
\newcommand{\Mob}{\mathop{\textrm{M\"ob}}}
\newcommand{\Aut}{\mathop{\mathrm{Aut}}}
\newcommand{\g}{\mathfrak{g}}
\renewcommand{\geq}{\geqslant}
\renewcommand{\leq}{\leqslant}
\newcommand{\PBC}{\text{PBC}}

\newcommand{\mb}{\bm}
\newcommand{\ua}{\uparrow}
\newcommand{\da}{\downarrow}
\newcommand{\ra}{\rightarrow}
\newcommand{\la}{\leftarrow}
\newcommand{\mc}{\mathcal}                                                                                                                                                                                                                                                                                                                                                                                                                                                                                                                                                                                
\newcommand{\bs}{\boldsymbol}
\newcommand{\lra}{\leftrightarrow}
\newcommand{\nn}{\nonumber}
\newcommand{\half}{{\textstyle{\frac{1}{2}}}}
\newcommand{\mf}{\mathfrak}
\newcommand{\MF}{\text{MF}}
\newcommand{\IR}{\text{IR}}
\newcommand{\UV}{\text{UV}}

\renewcommand{\thetable}{S\arabic{table}}
\renewcommand{\thefigure}{S\arabic{figure}}
\renewcommand{\thesection}{S\arabic{section}}
\renewcommand{\thesubsection}{\arabic{subsection}}
\renewcommand{\theequation}{S\arabic{equation}}

\DeclareGraphicsExtensions{.png}

\title{Supplementary Information Appendix for\\
``Automorphic Bloch theorems for hyperbolic lattices''}

\author{Joseph Maciejko}
\email{maciejko@ualberta.ca}
\affiliation{Department of Physics \& Theoretical Physics Institute (TPI), University of Alberta, Edmonton, Alberta T6G 2E1, Canada}
\author{Steven Rayan}
\email{rayan@math.usask.ca}
\affiliation{Department of Mathematics and Statistics\\\& Centre for Quantum Topology and Its Applications (quanTA), University of Saskatchewan, Saskatoon, Saskatchewan S7N 5E6, Canada}

\maketitle


\section{Computing the transversal}
\label{app:transversal}

As Eq.~(12) in the main text makes clear, the transversal $T$ is the set of translations $g_i$ that tile $\c{C}$ with $N$ copies of $\c{D}$. To obtain a connected cluster of given size $N$, we proceed as in Ref.~\cite{boettcher2021}, searching for group elements $g_i$ of increasing word length $L$ in the generators $\gamma_j$ of $\Gamma$, where $L$ is defined as the smallest possible number of generators appearing in the product expressing $g_i$. The set of (inequivalent) words of given $L$ in $\Gamma$, at least for small $L$, can be obtained by brute-force computational enumeration and elimination of redundancies using the $PSU(1,1)$ representation of the $\gamma_j$~\cite{boettcher2021}. For $L$ up to 3, we have verified that the number of inequivalent words thus obtained matches the growth function of $\Gamma$, which can be computed directly in GAP (i.e., the number of distinct elements of smallest word length $L$). The unique word of length 0 is the identity $e=g_1$, which we canonically take to be the first element of $T$. The 8 words of length 1 are the four generators $\gamma_j$ and their inverses $\gamma_j^{-1}$, which when acting on $\c{D}$, produce 8 octagons adjacent to the 8 sides of $\c{D}$. The 56 words of length 2 form an inequivalent subset of the 64 possible products of two words of length 1. As the computational results presented in this work are limited to clusters of size $N\leq N_\text{max}=25$, we can choose to limit our search to connected clusters for which words of length 3 and longer do not appear in the transversal. Though smaller than the total number of possible PBC clusters of a given size $N$, our results indicate that such clusters can be found at every $N\leq N_\text{max}$, which is sufficient to illustrate our ideas. To summarize, the transversal for a cluster of size $N_\text{max}$ takes the form:
\begin{align}
T=\{e,\gamma_1,\ldots,\gamma_4,\gamma_1^{-1},\ldots,\gamma_4^{-1},g_9,\ldots,g_{N_\text{max}}\},
\end{align}
where $g_9,\ldots,g_{N_\text{max}}$ are length-2 words. For a cluster of size $10\leq N\leq N_\text{max}$, only $N-9$ words of length 2 are kept. For a cluster of size $2\leq N\leq 9$, only $N-1$ words of length 1 are kept. Finally, there is only one cluster of size $N=1$, $\c{C}=\c{D}$, corresponding to $\Gamma_\PBC=\Gamma$.


\section{Clusters of prime order}

The number $\mathrm{NSG}_p$ of normal subgroups of prime index $N=p$ obeys a simple relation, Eq.~(13) in the main text:
\begin{align}\label{NSGprime}
\mathrm{NSG}_p=\frac{p^r-1}{p-1}=1+p+\ldots+p^{r-1},
\end{align}
where $r=4$. As Fig.~3 in the main text illustrates (compare blue circles and green line), this relation does not hold for nonprime values of $N$.

Relation (\ref{NSGprime}) can be derived as follows. A normal subgroup $\Gamma_\text{PBC}\triangleleft\Gamma$ of prime index $p$ is the kernel of the quotient homomorphism $f:\Gamma\rightarrow\Gamma/\Gamma_\text{PBC}$, where $\Gamma/\Gamma_\text{PBC}$ is a finite group of order $p$. Since $p$ is prime, $\Gamma/\Gamma_\text{PBC}$ is isomorphic to the cyclic group $\mathbb{Z}_p$, the additive group of integers mod $p$. Now $f$ is surjective, thus the number of normal subgroups of prime index $p$ corresponds to the number of distinct surjective homomorphisms $f:\Gamma\rightarrow\mathbb{Z}_p$. More precisely, two such maps that differ by composition with an automorphism of $\mathbb{Z}_p$ have the same kernel, thus we need to divide this number by the number $p-1$ of automorphisms of $\mathbb{Z}_p$. Since $\mathbb{Z}_p$ is abelian, $f$ factors through the abelianization $\Gamma\rightarrow\Gamma/[\Gamma,\Gamma]\cong\mathbb{Z}^r$ of $\Gamma$, where $r=2g=4$ for the Bolza lattice. There are $p^r$ homomorphisms from $\mathbb{Z}^r$ to $\mathbb{Z}_p$, obtained by assigning one of the $p$ elements of $\mathbb{Z}_p$ to each of the $r$ linearly independent generators of $\mathbb{Z}^r$. For $p$ prime, this homomorphism is surjective unless each generator is mapped to the identity in $\mathbb{Z}_p$. Thus there are $p^r-1$ surjective homomorphisms $f$. As said before, $\mathbb{Z}_p$ has $p-1$ automorphisms. Thus the number of distinct normal subgroups of $\Gamma$ is $(p^r-1)/(p-1)$.

An interesting physical consequence of this analysis is that PBC clusters with a prime number $p$ of unit cells are necessarily abelian, with a residual translation group isomorphic to $\mathbb{Z}_p$. Since by Euclid's theorem, prime numbers form an infinite sequence, this produces an infinite sequence of ``prime PBC clusters'' for which $U(1)$ hyperbolic band theory holds exactly.

\section{Translation matrices}

We now show that the translation matrices, Eq.~(26) in the main text, form a faithful representation of $\Gamma/\Gamma_\PBC$ and commute with the hopping matrix $H$. Using the group multiplication law~\footnote{The set of cosets of a {\it normal} subgroup $H$ in a group $G$ indeed form a group, since for any $g,g'\in G$, we have $[g][g']=(Hg)(Hg')=H(gHg^{-1})gg'=Hgg'=[gg']$.} in $\Gamma/\Gamma_\PBC$, i.e., $[g][g']=[gg']$, it is easily checked that Eq.~(26) forms a representation of this group,
\begin{align}
\Bigl(\c{U}([g_k])\c{U}([g_{k'}])\Bigr)_{i\ell}&=\sum_j \c{U}_{ij}([g_k])\c{U}_{j\ell}([g_{k'}])\nn\\
&=\sum_j\delta_{[g_i],[g_kg_j]}\delta_{[g_j],[g_{k'}g_\ell]}\nn\\
&=\delta_{[g_i],[g_kg_{k'}g_\ell]}\nn\\
&=\c{U}_{i\ell}([g_k g_{k'}]),
\end{align}
for any two $[g_k],[g_{k'}]\in\Gamma/\Gamma_\PBC$. As pointed out in {\it Nonabelian clusters: a nonabelian Bloch theorem} in the main text, those matrices form the regular representation of $\Gamma/\Gamma_\PBC$. The regular representation is a faithful representation.

Finally, translation symmetry on the finite cluster is the statement that the translation operators $\c{T}_{[g_k]}$ commute with $\c{H}_\PBC$ for all $[g_k]\in\Gamma/\Gamma_\PBC$. For this to hold, the translation matrices $\c{U}([g_k])$ must commute with the hopping matrix $H$. Noting that Eq.~(26) is a permutation matrix, and thus orthogonal, we have
\begin{align}
\Bigl(\c{U}^{-1}([g_k])H\c{U}([g_k])\Bigr)_{im}&=\sum_{j\ell}\c{U}_{ji}([g_k])H_{j\ell}\c{U}_{\ell m}([g_k])\nn\\
&=-\sum_\alpha\sum_{j\ell}\delta_{[g_j],[g_kg_i]}\delta_{[g_\ell],[g_j\gamma_\alpha]}\delta_{[g_\ell],[g_kg_m]}\nn\\
&=-\sum_\alpha\delta_{[g_kg_m],[g_kg_i\gamma_\alpha]},
\end{align}
using the explicit form Eq.~(24). The summand is nonzero if and only if $[g_kg_m]=[g_kg_i\gamma_\alpha]$, that is, if
\begin{align}
\Gamma_\text{PBC}g_kg_m=\Gamma_\text{PBC}g_kg_i\gamma_\alpha.
\end{align}
Multiplying on the left by $g_k^{-1}$ and using the normality of $\Gamma_\text{PBC}$ in $\Gamma$, we have $\Gamma_\text{PBC}g_m=\Gamma_\text{PBC}g_i\gamma_\alpha$, and thus
\begin{align}
\Bigl(\c{U}^{-1}([g_k])H\c{U}([g_k])\Bigr)_{im}=-\sum_\alpha\delta_{[g_m],[g_i\gamma_\alpha]}=H_{im}.
\end{align}


\begin{figure}[t]
\includegraphics[width=0.8\columnwidth]{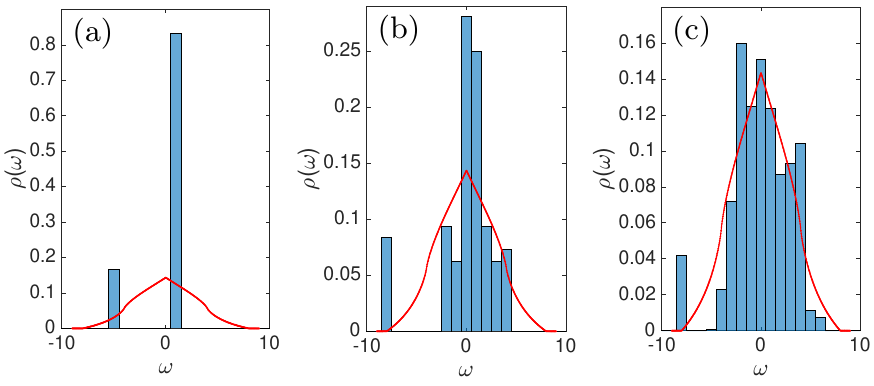}
 \caption{Density of states histograms (blue) for abelian PBC clusters of size (a) $N=6$, (b) $N=12$, and (c) $N=24$, compared with the exact density of states of a tight-binding model with unit nearest-neighbor amplitude on an infinite four-dimensional hypercubic lattice (red).}
  \label{fig:abelianDOS}
\end{figure}

\section{Density-of-states histograms and fraction of abelian states}

\subsection{Abelian clusters}

To get a sense of how well $\Jac(\Sigma_2)$ is sampled upon increasing the PBC cluster size $N$, we plot in Fig.~\ref{fig:abelianDOS} density-of-states (DOS) histograms for three different cluster sizes. Here we present histogram data averaged over all possible abelian clusters of a given size, to reduce finite-size effects. The finite-size histogram data can be compared with the expected DOS for an infinite abelian cluster, i.e., for the spectrum (30) in the main text but with $\b{k}\in\Jac(\Sigma_2)$ a continuous variable. This corresponds to the DOS for a tight-binding model with unit nearest-neighbor hopping amplitude on the four-dimensional hypercubic lattice:
\begin{align}\label{4dDOS1}
\rho_\text{4D}(\omega)=\int_{\Jac(\Sigma_2)}\frac{d^4k}{(2\pi)^4}\delta(\omega-E(\b{k})).
\end{align}
This can be expressed in terms of the DOS for a nearest-neighbor tight-binding model on the simple cubic lattice,
\begin{align}\label{4dDOS}
\rho_\text{4D}(\omega)=\int_{-\pi}^\pi\frac{dk_4}{2\pi}\rho_\text{3D}(\omega+2\cos k_4),
\end{align}
which can be computed analytically:
\begin{align}
\rho_\text{3D}(\omega)=-\frac{1}{\pi}\Im\left[\frac{1}{\omega+i\eta}P\left(\frac{6}{\omega+i\eta}\right)\right],
\end{align}
where $\eta$ is a positive infinitesimal, and we define~\cite{joyce1972}:
\begin{figure}[t]
\includegraphics[width=0.6\columnwidth]{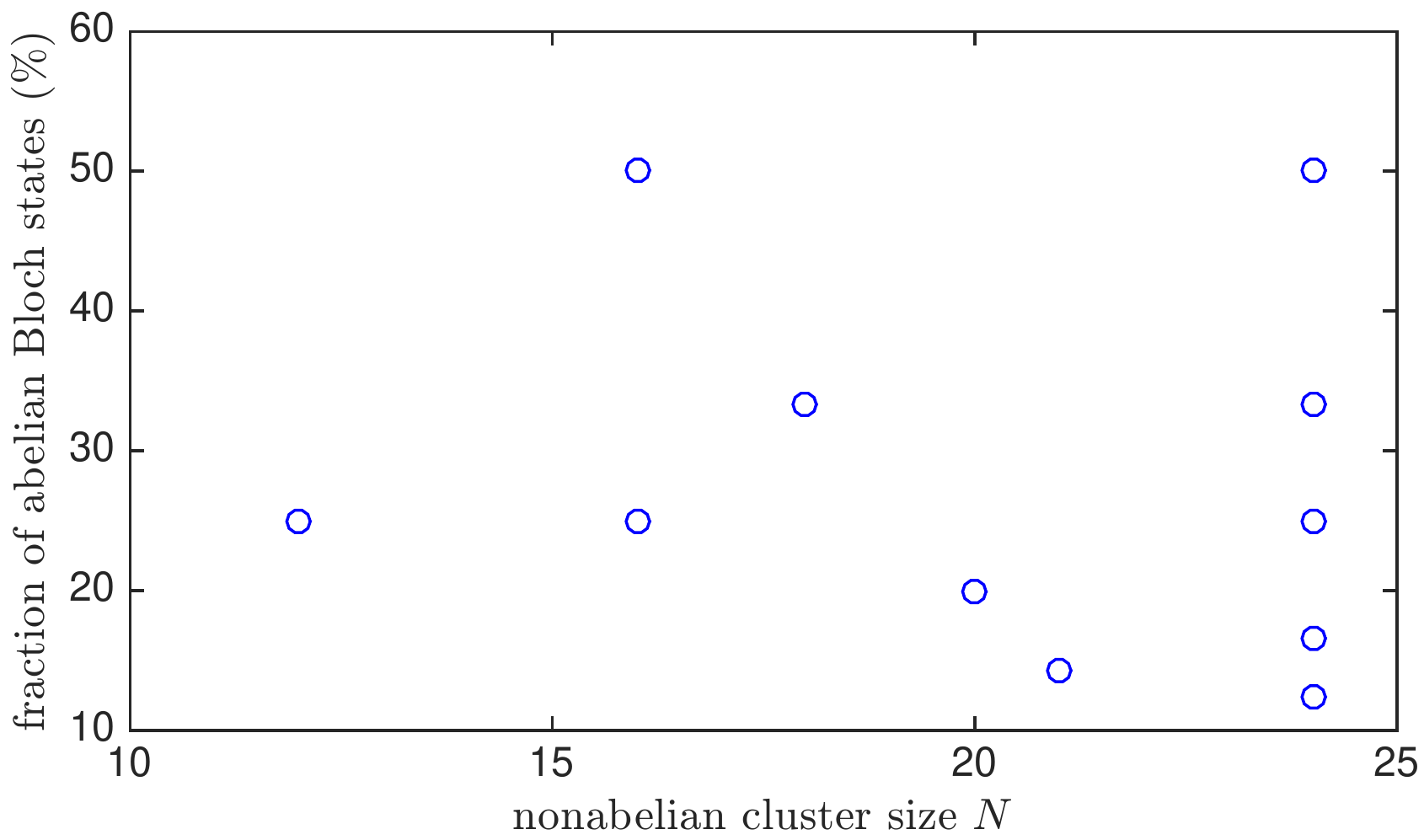}
\caption{Fraction $N_\text{ab}/N$ of eigenstates obeying the abelian Bloch theorem [Eq.~(27)] relative to all eigenstates, for nonabelian PBC clusters of size $N\in\{12,16,18,20,21,24\}$.}
\label{fig:irrepfrac}
\end{figure}
\begin{align}
P(y)&=\frac{\sqrt{1-\frac{3}{4}x_1}}{1-x_1}\left(\frac{2}{\pi}\right)^2 K(k_+^2)K(k_-^2),\\
k_\pm^2&=\frac{1}{2}\pm\frac{1}{4}x_2\sqrt{4-x_2}-\frac{1}{4}(2-x_2)\sqrt{1-x_2},\\
x_2&=\frac{x_1}{x_1-1},\\
x_1&=\frac{1}{2}+\frac{1}{6}y^2-\frac{1}{2}\sqrt{(1-y^2)\left(1-\frac{1}{9}y^2\right)},
\end{align}
and $K(m)$ is the complete elliptic integral of the first kind:
\begin{align}
K(m)=\int_0^{\pi/2}\frac{d\theta}{\sqrt{1-m\sin^2\theta}}.
\end{align}
The remaining integral over $k_4$ in Eq.~(\ref{4dDOS}) can be performed numerically. It is clear from Eqs.~(30) and (\ref{4dDOS1}) that the DOS vanishes for $\omega$ outside the interval $[-8,8]$. We see from Fig.~\ref{fig:abelianDOS} that as $N$ increases, the finite-size DOS histograms approximate the exact DOS increasingly well.

\begin{figure}[t]
\includegraphics[width=0.8\columnwidth]{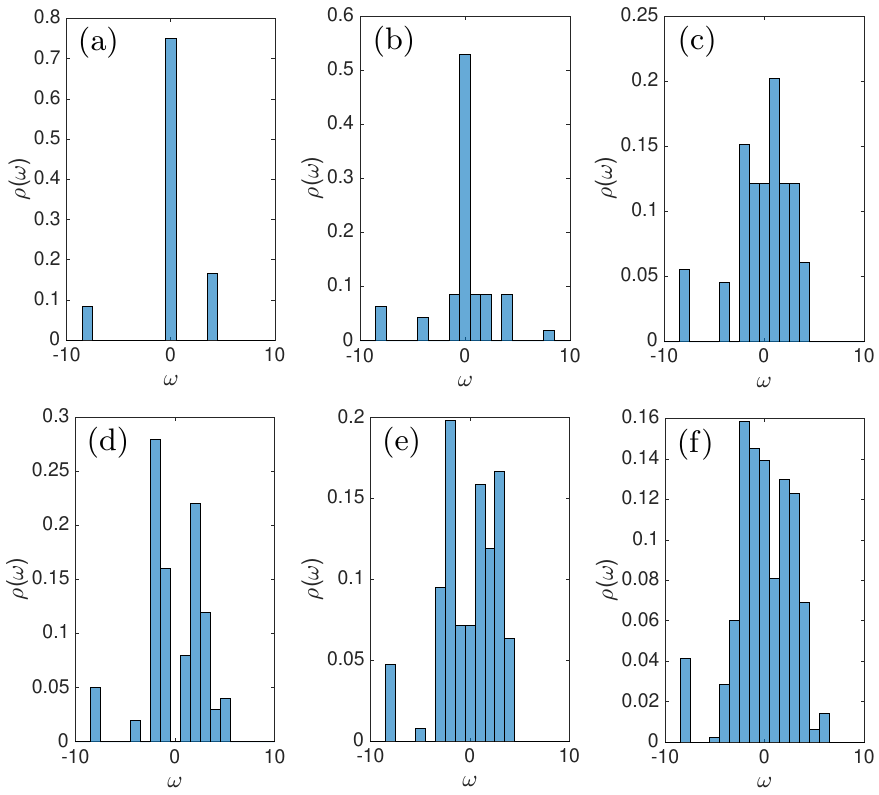}
 \caption{Density of states histograms for nonabelian PBC clusters of size (a) $N=12$, (b) $N=16$, (c) $N=18$, (d) $N=20$, (e) $N=21$, and (f) $N=24$.}
  \label{fig:nonabelianDOS}
\end{figure}

\subsection{Nonabelian clusters}

For a given nonabelian cluster, the fraction $N_\text{ab}/N$ of abelian states among all eigenstates can be determined from the number $\c{N}_\text{1D}$ of one-dimensional irreps of $\Gamma/\Gamma_\PBC$, which can be computed using representation-theoretic routines in GAP. Indeed, since each one-dimensional irrep occurs only once in the direct-sum decomposition (32) in the main text, we have $N_\text{ab}=\c{N}_\text{1D}$. In Fig.~\ref{fig:irrepfrac}, we plot for each cluster size $N\in\{12,16,18,20,21,24\}$ the distinct values of $N_\text{ab}/N$ found across all possible nonabelian clusters of size $N$. As the cluster size increases, the minimum value of $N_\text{ab}/N$ appears to decrease, which suggests higher-dimensional representations play an increasingly important role for larger clusters. However, for the largest system size considered, nonabelian clusters can nonetheless still be found for which up to 50\% of the spectrum consists of abelian states. Interestingly, the averaged DOS histograms for nonabelian clusters (Fig.~\ref{fig:nonabelianDOS}) are qualitatively not too dissimilar from those for abelian clusters (Fig.~\ref{fig:abelianDOS}), despite the presence of a substantial fraction of nonabelian states. The reason for this unexpected similarity is not presently understood and deserves further investigation.

The observations above, combined with the fact that nonabelian clusters only appear at certain values of $N$, lead us to conjecture that nonabelian clusters are those that exhibit a higher degree of symmetry than their abelian counterparts. In exact diagonalization studies of quantum Hamiltonians on ordinary Euclidean lattices, certain high-symmetry points in the Brillouin zone will be excluded by an anisotropic choice of cluster geometry. Likewise here, the relative fraction of abelian to nonabelian ``Brillouin zones'' appearing in the spectrum can be tuned by changing the geometry of the cluster, to the point of completely excluding nonabelian representations for certain system sizes.

\section{Irrep decomposition for a nonabelian cluster}

\begin{table}[t]
\begin{tabular}{|c||c|c|c|c|c|c|c|c|c|c|c|c|}
\hline
$C$ & $1$ & $2$ & $3$ & $4$ & $5$ & $6$ & $7$ & $8$ & $9$ & $10$ & $11$ & $12$  \\ \hline
$n_C$ & $1$ & $1$ & $1$ & $1$ & $2$ & $2$ & $2$ & $2$ & $3$ & $3$ & $3$ & $3$  \\
\hline\hline
$D^{(1)}$ & $1$ & $1$ & $1$ & $1$ & $1$ & $1$ & $1$ & $1$ & $1$ & $1$ & $1$ & $1$ \\ \hline
$D^{(2)}$ & $1$ & $1$ & $1$ & $1$ & $1$ & $1$ & $1$ & $1$ & $-1$ & $-1$ & $-1$ & $-1$ \\ \hline
$D^{(3)}$ & $1$ & $-1$ & $a$ & $-a$ & $1$ & $-1$ & $a$ & $-a$ & $c$ & $-c$ & $-1/c$ & $1/c$ \\ \hline
$D^{(4)}$ & $1$ & $-1$ & $a$ & $-a$ & $1$ & $-1$ & $a$ & $-a$ & $-c$ & $c$ & $1/c$ & $-1/c$ \\ \hline
$D^{(5)}$ & $1$ & $-1$ & $-a$ & $a$ & $1$ & $-1$ & $-a$ & $a$ & $-1/c$ & $1/c$ & $c$ & $-c$ \\ \hline
$D^{(6)}$ & $1$ & $-1$ & $-a$ & $a$ & $1$ & $-1$ & $-a$ & $a$ & $1/c$ & $-1/c$ & $-c$ & $c$ \\ \hline
$D^{(7)}$ & $1$ & $1$ & $-1$ & $-1$ & $1$ & $1$ & $-1$ & $-1$ & $a$ & $a$ & $-a$ & $-a$ \\ \hline
$D^{(8)}$ & $1$ & $1$ & $-1$ & $-1$ & $1$ & $1$ & $-1$ & $-1$ & $-a$ & $-a$ & $a$ & $a$ \\ \hline
$D^{(9)}$ & $2$ & $2$ & $-2$ & $-2$ & $-1$ & $-1$ & $1$ & $1$ & $0$ & $0$ & $0$ & $0$ \\ \hline
$D^{(10)}$ & $2$ & $2$ & $2$ & $2$ & $-1$ & $-1$ & $-1$ & $-1$ & $0$ & $0$ & $0$ & $0$ \\ \hline
$D^{(11)}$ & $2$ & $-2$ & $b$ & $-b$ & $-1$ & $1$ & $-a$ & $a$ & $0$ & $0$ & $0$ & $0$ \\ \hline
$D^{(12)}$ & $2$ & $-2$ & $-b$ & $b$ & $-1$ & $1$ & $a$ & $-a$ & $0$ & $0$ & $0$ & $0$ \\ \hline
\end{tabular}
\caption{Character table of $\Gamma/\Gamma_\PBC$ for the $N=24$ nonabelian cluster considered in {\it Irrep decomposition of the finite-size spectrum: an explicit example}. Here $a=-i$, $b=-2i$, and $c=e^{-i\pi/4}$. $C$ denotes the conjugacy class, $n_C$ the number of group elements in each class, and $D^{(\lambda)}$ the 12 irreps labeled by $\lambda=1,\ldots,12$.}
\label{tab:char}
\end{table}

For the chosen nonabelian PBC cluster of size $N=24$, the irreducible character table of the finite group $\Gamma/\Gamma_\PBC$ of order 24, computed by GAP, is shown in Table~\ref{tab:char}. The number of conjugacy classes/irreps is found to be $\c{N}=12$.  The first row of Table~\ref{tab:char} denotes the label $C=1,\ldots,\c{N}$ of the conjugacy class, and the second, the number $n_C$ of group elements in each class. One can check explicitly that the characters $\chi^{(\lambda)}=\tr D^{(\lambda)}$ of the irreps $\lambda=1,\ldots,\c{N}$, which depend only on the class $C$, satisfy the properties of row orthogonality,
\begin{align}
\sum_{C=1}^\c{N}n_C\chi^{(\lambda)}(C)^*\chi^{(\lambda')}(C)=N\delta_{\lambda\lambda'},
\end{align}
and column orthogonality,
\begin{align}
\sum_{\lambda=1}^\c{N}\chi^{(\lambda)}(C)^*\chi^{(\lambda)}(C')=\frac{N}{n_C}\delta_{CC'},
\end{align}
where $N=24$ is the order of the group~\cite{Tinkham}. The first column of the table corresponds to the dimension $r_\lambda=\chi^{(\lambda)}(e)$ of irrep $\lambda$; thus this group has 8 one-dimensional irreps $\lambda=1,\ldots,8$ and 4 two-dimensional irreps $\lambda=9,\ldots,12$. 

To determine to which irrep each Bloch eigenstate of the cluster belongs, we construct projector matrices~\cite{Tinkham},
\begin{align}
\Pi^{(\lambda)}=\frac{r_\lambda}{N}\sum_{[g_k]\in\Gamma/\Gamma_\PBC}\chi^{(\lambda)}([g_k])^*\c{U}([g_k]),
\end{align}
which obey $\Pi^{(\lambda)}\Pi^{(\lambda')}=\delta_{\lambda\lambda'}\Pi^{(\lambda)}$ and project an arbitrary state $\psi(z_i)$ onto irrep $\lambda$. Since the $\Pi^{(\lambda)}$ are linear combinations of translation matrices $\c{U}$, they commute with $H$ and can thus be simultaneously diagonalized with it. Finally, we note that the eigenvalues of $\Pi^{(\lambda)}$ can only be $1$ or $0$, and that an eigenstate of $H$ can have a $\Pi^{(\lambda)}$-eigenvalue of $1$ for only a single $\Pi^{(\lambda)}$. Such an eigenstate $\psi^{(\lambda)}_\mu$ thus necessarily belongs to irrep $\lambda$ of $\Gamma/\Gamma_\PBC$.


\section{Geometry of higher-rank moduli spaces}

In this section, we provide further information on the geometry of the rank-$r$ moduli spaces for general $r$. While a positively-curved $SU(r)$ fibre is a feature of these moduli spaces in general—in algebro-geometric language, $\c M(\Sigma_g,SU(r))$ is \emph{Fano}, as opposed to tori which are \emph{Calabi--Yau}---we also caution that the $g = r = 2$ example is rather unusual in that (a) the moduli space happens to be globally smooth and (b) the geometry of the moduli space is known in an exact way.  If we let $\c{M}(\Sigma_g,U(r),d)$ denote the moduli space of semistable rank-$r$ holomorphic vector bundles with first Chern class $c_1(V)=d$, then it is worth noting that the moduli space is only guaranteed to be smooth when $r$ and $d$ are coprime~\cite{narasimhan1965}.  In the coprime case, it is impossible to achieve equality in the stability inequality, and so the potentially singular points disappear from the moduli space.  In our case of $\c{M}(\Sigma_g,U(r))=\c{M}(\Sigma_g,U(r),0)$, this coprimality never occurs and we do have bundles for which equality is achieved, which makes the global smoothness of $\c{M}(\Sigma_2,U(2))$ somewhat surprising.  In terms of $\pi_1(\Sigma_g)$-representations, coprimality of $r$ and $d$ means that we may construct a compact character variety from irreps alone; however, the nonzero first Chern class must be incorporated and manifests as a twist by a nontrivial root of unity in the defining relation of the character variety.  For our purposes, we will restrict to the $d=0$ case, in which case we have the ordinary character variety as the rank-$r$ component of the space of crystal momenta.

Physically, one can still prescribe meaning to $\c{M}(\Sigma_g,U(r))$ as a space of higher-dimensional fluxes, but the (co)homologies of the $\c{M}(\Sigma_g,U(r))$---in particular the relations that intertwine cycles---are far more involved now than that of $\Jac(\Sigma_g)$.   We note, however, that there is a natural map $\c{M}(\Sigma_g,U(r))\to\Jac(\Sigma_g)$ obtained by $V\mapsto\det(V)$, where $\det(V)$ is the line bundle whose transition functions are the determinants of those of $V$.  In other words, there is an algebraic projection of higher-dimensional fluxes onto the magnetic fluxes in the abelian Brillouin zone.


\section{Experimental proposal}

\begin{figure}[t]
\includegraphics[width=\columnwidth]{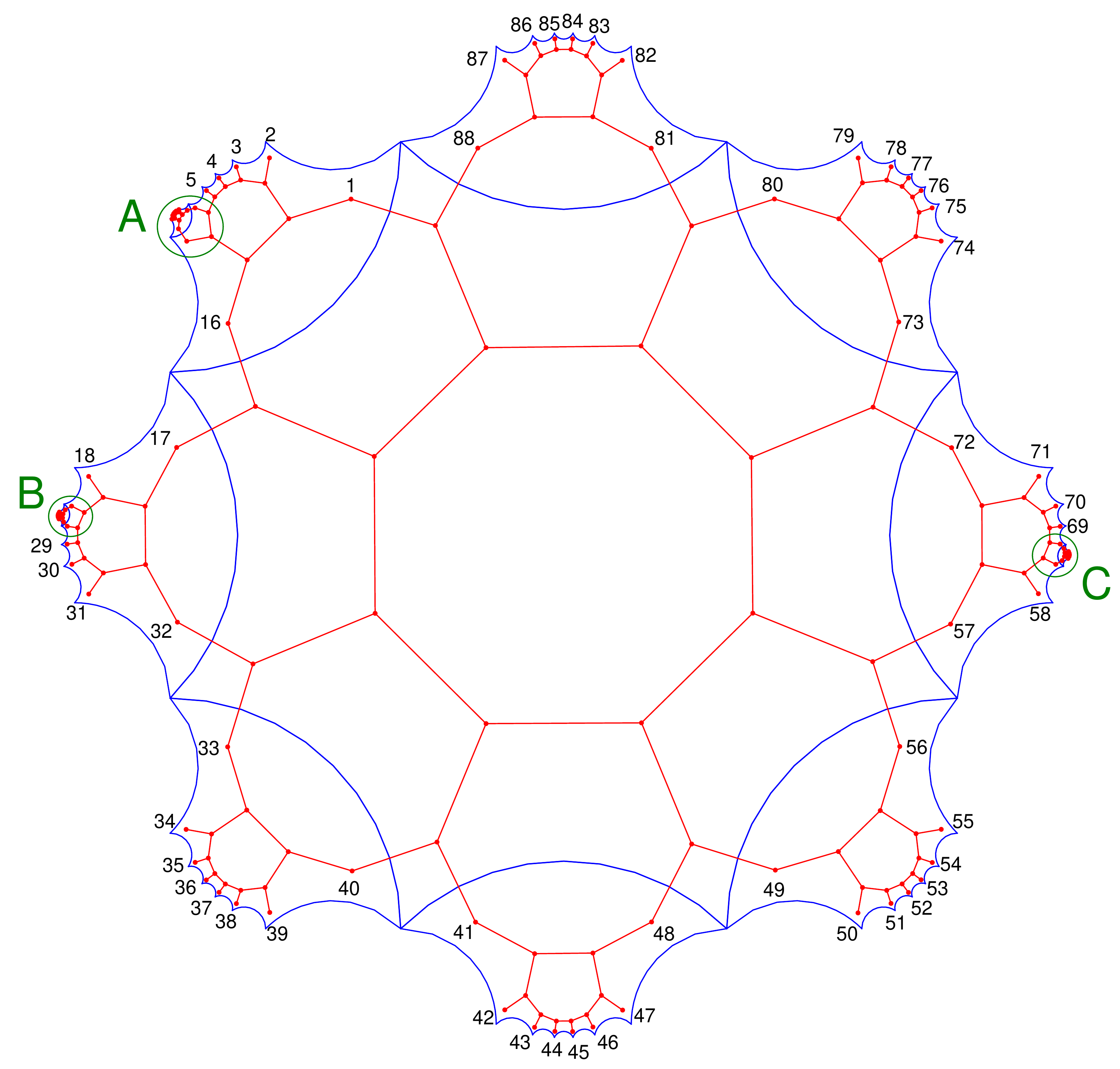}
\caption{Proposed realization of abelian/nonabelian Bloch states using CQED or electric circuit implementations (enlarged version of Fig.~9 in the main text). A cluster of 192 sites (red vertices) of the $\{8,3\}$ lattice corresponds to $N=12$ Bolza unit cells (blue octagons) with 16 sites each. Close-ups of the A, B, C boundary octagons (green) are given in Fig.~\ref{fig:k271ABC}. The 88 boundary sites can be wired together in two different ways to achieve either abelian PBC (Table~\ref{tab:k271}) or nonabelian PBC (Table~\ref{tab:k281}).}
\label{fig:k271}
\end{figure}

\begin{figure}[t]
\centering\includegraphics[width=0.54\columnwidth]{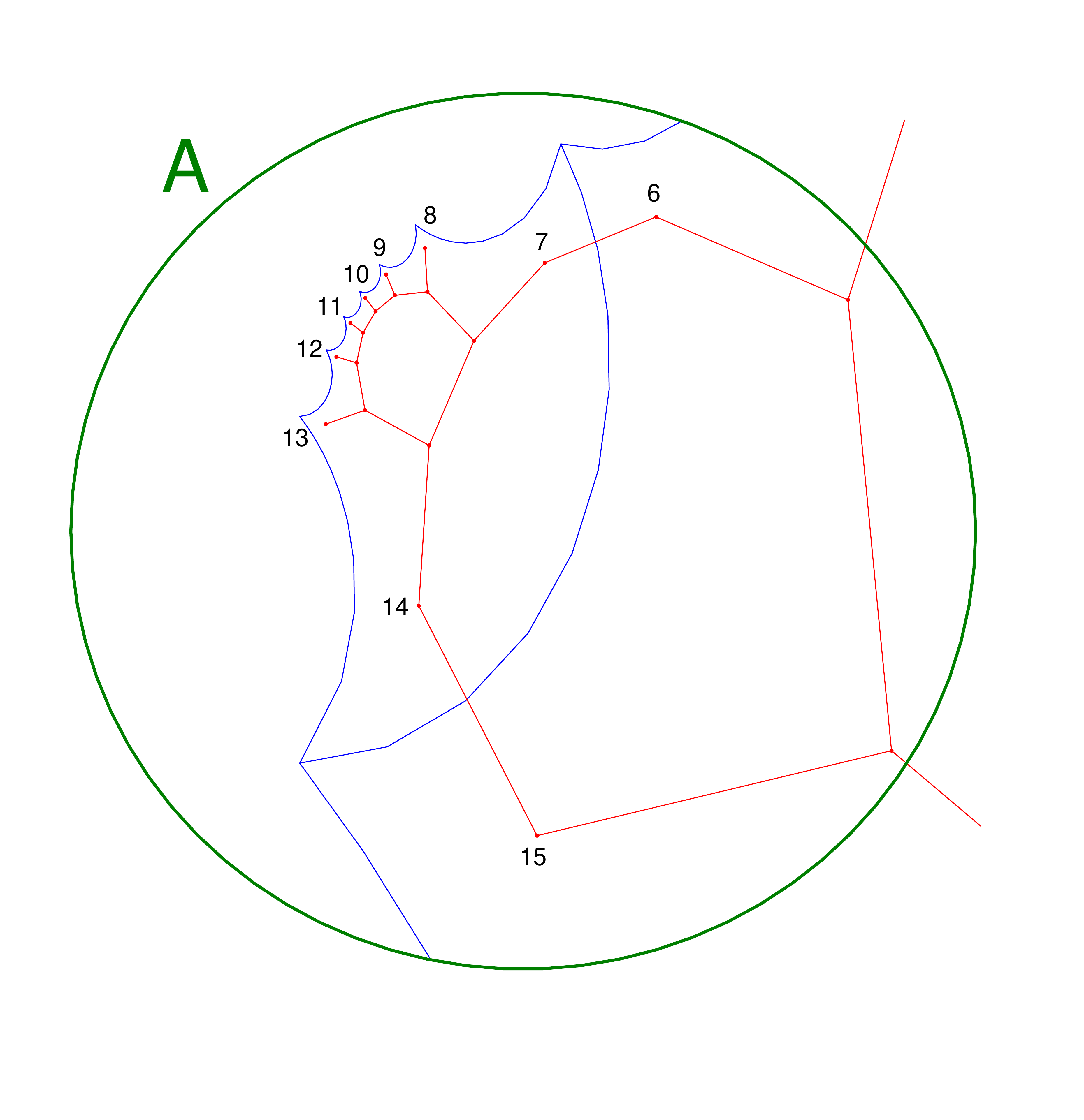}
\centering\raisebox{0.14\height}{\includegraphics[width=0.45\columnwidth]{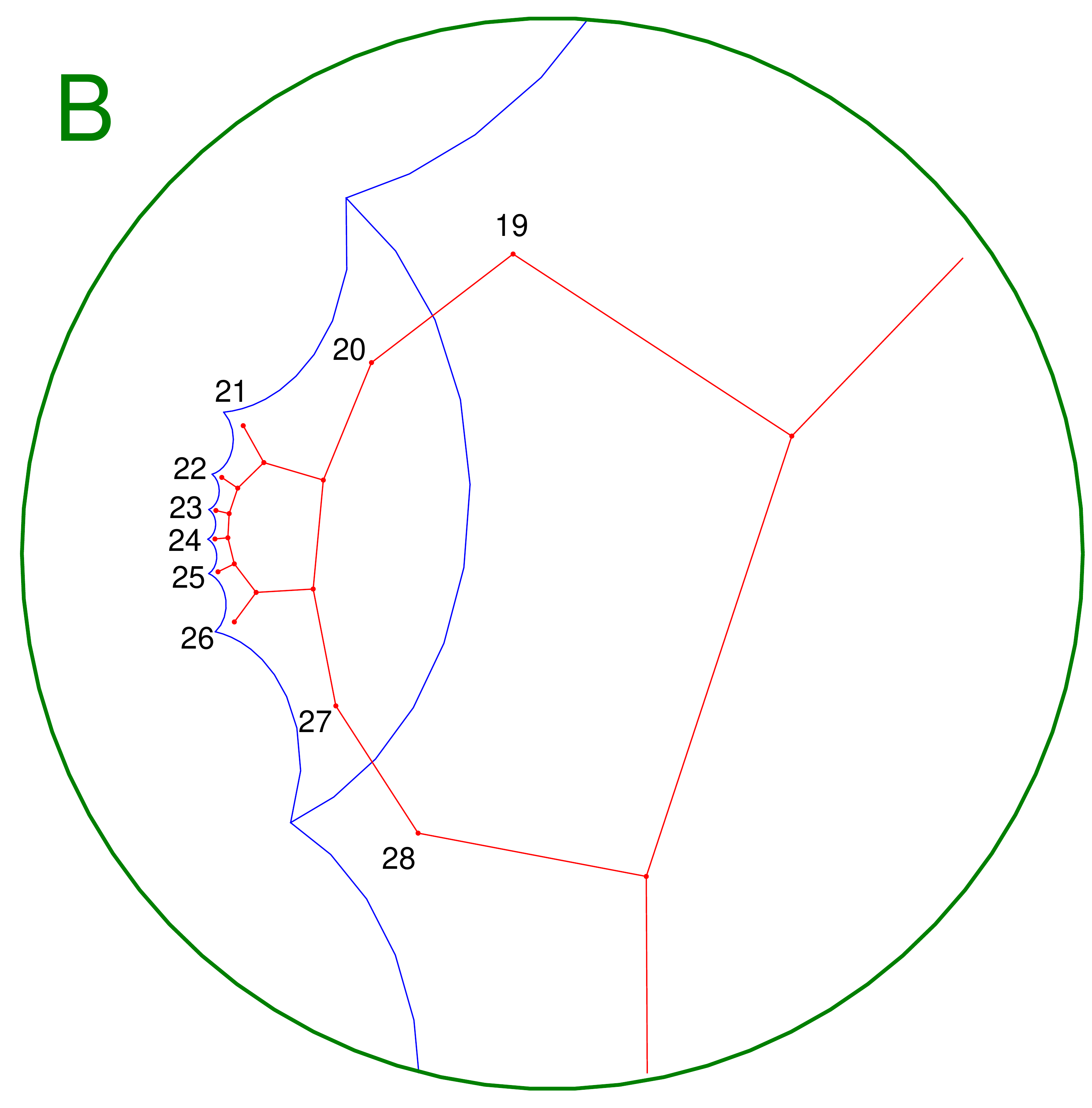}}
\centering\includegraphics[width=0.55\columnwidth]{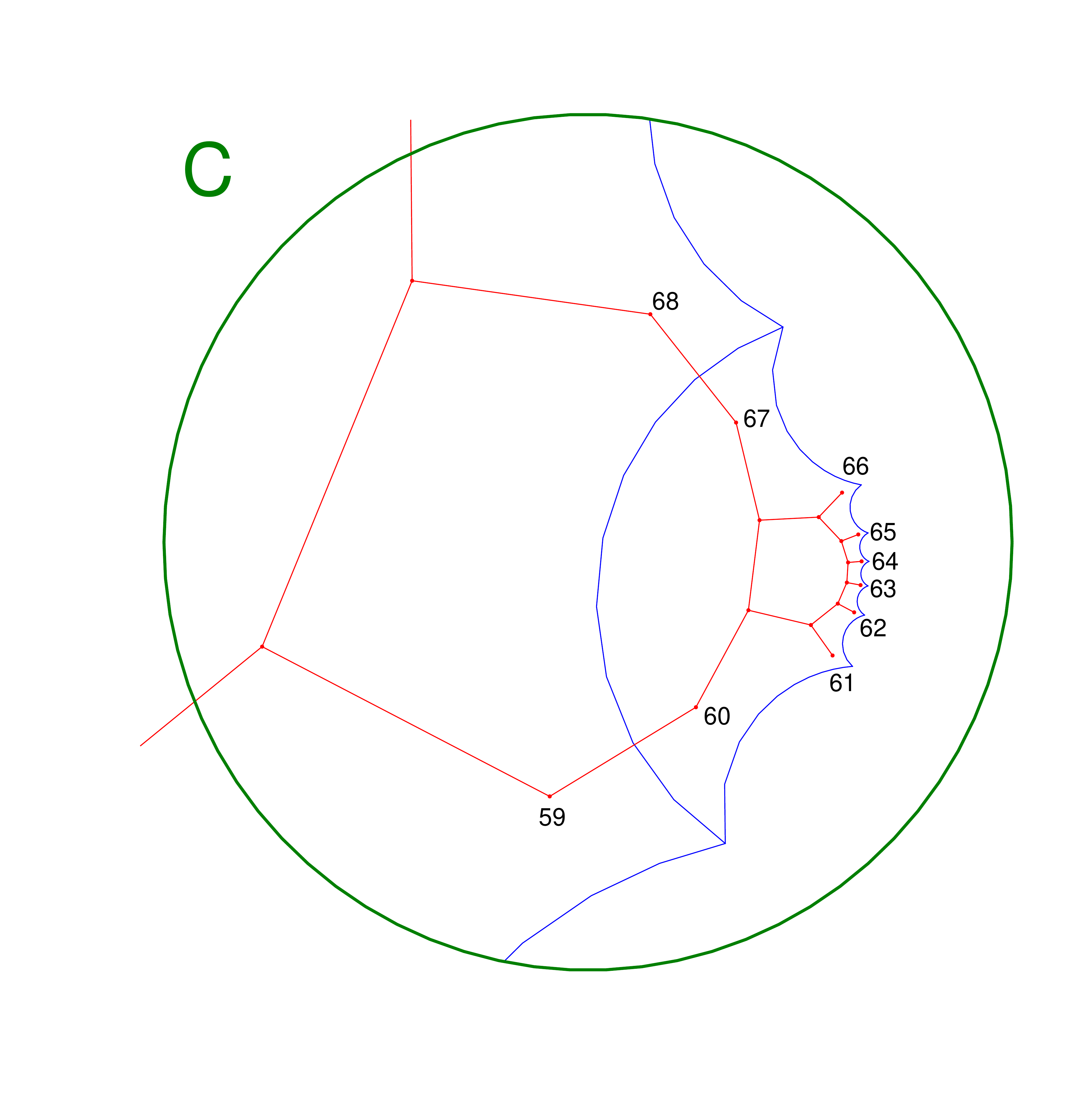}
\caption{Close-ups of the A, B, C boundary octagons in Fig.~\ref{fig:k271}.}
\label{fig:k271ABC}
\end{figure}

In this last section, we provide further details of the proposal outlined in {\it Summary and outlook} to realize hyperbolic abelian/nonabelian Bloch states in CQED and/or electric circuit network experiments. Figure~9 in the main text is reproduced in enlarged format as Fig.~\ref{fig:k271}; close-ups of the boundary octagons A, B, C (green) are given in Fig.~\ref{fig:k271ABC}.

To engineer PBC that are consistent with the normal subgroup structure discussed in the main text, the 88 boundary sites must be wired together in particular ways. In Tables \ref{tab:k271} and \ref{tab:k281}, we list the 74 distinct bonds between boundary sites $i$ and $j>i$ that must be established to engineer (a particular choice of) abelian and nonabelian PBC, respectively. This information is also included in the SI files Dataset 1 and Dataset 2, respectively, where the first (second) column corresponds to the value of $i$ ($j$). The hopping strength (capacitive coupling) must be the same for those boundary bonds as for the intra-cluster bonds (red nearest-neighbor bonds in Figs.~\ref{fig:k271} and \ref{fig:k271ABC}) in order to preserve Fuchsian translation symmetry and the automorphic Bloch theorem. With either abelian or nonabelian PBC, both bulk and boundary sites are connected to 3 nearest neighbors (as can be checked from Tables~\ref{tab:k271}-\ref{tab:k281}), again consistent with translation symmetry on the $\{8,3\}$ lattice.

\clearpage

\clearpage

\begin{table}[t]
\begin{tabular}{|c|c||c|c||c|c||c|c|}
\hline
	$i$	&	$j$		&	$i$ & $j$	&	$i$ & $j$	&	$i$ & $j$	\\ \hline\hline
     1  &  36		&	14 & 86		&	29 & 68		&	44 & 71,76	\\ \hline
     2  &  30,35		&	15 & 67		&	30 & 85		&	45 & 70		\\ \hline
     3  &  29,52		&	16 & 66		&	31 & 79,84	&	49 & 76		\\ \hline
     4  &  46,51		&	17 & 62		&	32 & 78		&	50 & 70,75	\\ \hline
     5  &  21,45		&	18 & 61		&	33 & 63		&	51 & 69		\\ \hline
     6  &  20		&	21 & 58		&	34 & 62		&	52 & 86		\\ \hline
     7  &  59		&	22 & 57,74	&	36 & 84		&	53 & 61,85	\\ \hline
     8  &  58,75		&	23 & 73,82	&	37 & 78,83	&	54 & 60		\\ \hline
     9  &  47,74		&	24 & 40,81	&	38 & 72,77	&	64 & 80		\\ \hline
    10  &  46,55		&	25 & 39,48	&	39 & 71		&	65 & 79,88	\\ \hline
    11  &  19,54		&	26 & 47,56	&	41 & 64		&	66 & 87		\\ \hline
    12  &  18,35		&	27 & 55		&	42 & 63,83	&		&		\\ \hline
    13  &  34,87		&	28 & 69		&	43 & 77,82	&		&		\\ \hline
\end{tabular}
\caption{List of connections among boundary sites $i$ and $j$ of $\{8,3\}$ cluster of Fig.~\ref{fig:k271} to engineer abelian PBC (data also included in Dataset 1).}
\label{tab:k271}
\end{table}

\begin{table}[t]
\begin{tabular}{|c|c||c|c||c|c||c|c|}
\hline
	$i$	&	$j$		&	$i$ & $j$	&	$i$ & $j$	&	$i$ & $j$	\\ \hline\hline
     1  &  43		&	14 & 86		&	28 & 69		&	45 & 79,84	\\ \hline
     2  &  30,42		&	15 & 23		&	29 & 68		&	46 & 61,78	\\ \hline
     3  &  29,38		&	16 & 22		&	30 & 78		&	47 & 60		\\ \hline
     4  &  37,53		&	18 & 63		&	31 & 77,86	&	49 & 76		\\ \hline
     5  &  52,65		&	19 & 62		&	32 & 85		&	50 & 75,84	\\ \hline
     6  &  64		&	20 & 80		&	34 & 64		&	51 & 71,83	\\ \hline
     7  &  59		&	21 & 58,79	&	35 & 63,83	&	52 & 70		\\ \hline
     8  &  58,75		&	22 & 57		&	36 & 77,82	&	54 & 62		\\ \hline
     9  &  40,74		&	23 & 82		&	37 & 76		&	55 & 61		\\ \hline
    10  &  18,39		&	24 & 54,81	&	38 & 72		&	65 & 88		\\ \hline
    11  &  17,56		&	25 & 48,53	&	39 & 71		&	66 & 74,87	\\ \hline
    12  &  42,55		&	26 & 34,47	&	43 & 70		&	67 & 73		\\ \hline
    13  &  41,87		&	27 & 33		&	44 & 69,85	&	   &			\\ \hline
\end{tabular}
\caption{List of connections among boundary sites $i$ and $j$ of $\{8,3\}$ cluster of Fig.~\ref{fig:k271} to engineer nonabelian PBC (data also included in Dataset 2).}
\label{tab:k281}
\end{table}

\clearpage

\bibliography{PBC}